\documentclass[notoc]{JHEP3}
\usepackage{graphicx}
\usepackage{amsmath}
\usepackage{amsfonts}
\usepackage{amssymb}
\usepackage{epsfig}
\def \als {\alpha_{\mathrm{s}}}

\def \be {\mathbf{E}}

\def \br {\mathbf{r}}
\def \brg {\mathbf{R}}

\def \cf {C_F}

\def \nc {N_c}

\def \bk {\mathbf{k}}

\def \bp {\mathbf{p}}
\def \bpg {\mathbf{P}}

\newcommand{\Tint}[1]{{\hbox{$\sum$}\!\!\!\!\!\!\!\int\,}_{\!\!\!\!\raise-0.9ex\hbox{$\scriptstyle{#1}$}}}

\def \m2   {\mu^{2 \epsilon}}
\def\alVs{\alpha_{V_s}}
\def\alVo{\alpha_{V_o}}
\def\siml{{\ \lower-1.2pt\vbox{\hbox{\rlap{$<$}\lower6pt\vbox{\hbox{$\sim$}}}}\ }}
\def\simg{{\ \lower-1.2pt\vbox{\hbox{\rlap{$>$}\lower6pt\vbox{\hbox{$\sim$}}}}\ }}

\newcommand{\order}[1]{\mathcal{O}\left(#1\right)}

\title{Heavy Quarkonium in a weakly-coupled quark-gluon plasma
below the melting temperature}

\author{Nora Brambilla\\
Physik-Department, Technische Universit\"at M\"unchen,
James-Franck-Str. 1, 85748 Garching, Germany}
\author{Miguel \'Angel Escobedo\\
Departament d'Estructura i Constituents de la Mat\`eria and Institut de Ci\`encies del Cosmos, 
Universitat de Barcelona, Diagonal 647, E-08028, Barcelona, Catalonia, Spain}
\author{Jacopo Ghiglieri\\
Physik-Department, Technische Universit\"at M\"unchen,
James-Franck-Str. 1, 85748 Garching, Germany and Excellence Cluster Universe, Technische Universit\"at M\"unchen, 
Boltzmannstr. 2, 85748, Garching, Germany }
\author{Joan Soto\\
Departament d'Estructura i Constituents de la Mat\`eria and Institut de Ci\`encies del Cosmos, 
Universitat de Barcelona, Diagonal 647, E-08028, Barcelona, Catalonia, Spain}
\author{Antonio Vairo\\
Physik-Department, Technische Universit\"at M\"unchen,
James-Franck-Str. 1, 85748 Garching, Germany}

\preprint{TUM-EFT 7/10 \\ UB-ECM-PF-10/17 \\ ICCUB-10-033}

\abstract{We calculate the heavy quarkonium energy levels and decay
  widths in a quark-gluon plasma, whose temperature $T$ and screening
  mass $m_D$ satisfy the hierarchy $m\als \gg T \gg m \als^2 \gg m_D$
  ($m$ being the heavy-quark mass), at order $m\als^5$. We first
  sequentially integrate out the scales $m$, $m\als$ and $T$, and,
  next, we carry out the calculations in the resulting effective
  theory using techniques of integration by regions. A collinear
  region is identified, which contributes at this order. We also discuss
  the implications of our results concerning heavy quarkonium
  suppression in heavy ion collisions.}

\begin{document}

\section{Introduction}
Heavy quarkonium has been suggested since long time as a thermometer
for the medium that forms at the core of heavy-ion collision
experiments \cite{Matsui:1986dk}.  The early arguments were based on
the na\"\i ve expectation that above the deconfinement temperature the
confining part of the quark-antiquark potential vanishes and the
Coulomb part turns into a Yukawa potential due to screening. Since the
Yukawa potential supports a finite number of bound states depending on
the screening (Debye) mass, and the latter is linear in the
temperature, it is then clear that the relative fraction of the
different heavy quarkonium states observed will depend on the
temperature of the medium. In addition, the electromagnetic decays of
these states provide a clean experimental signature. The gross picture
above appears to be supported by experiments \cite{Lourenco:2006sr}.

In the last few years, significant progress has been made in 
deriving the quark-antiquark potential on a rigorous basis.
A model independent study of the real-time 
static potential was initiated for large temperatures ($T \gg 1/r \simg m_D$)
in \cite{Laine:2006ns,Laine:2007gj,Burnier:2007qm,Beraudo:2007ky}
and its implications for a QED and QCD plasma studied. 
For a wider range of temperatures, including lower temperatures, 
an effective field theory (EFT) study of non-relativistic bound states in a plasma 
was initiated in \cite{Escobedo:2008sy} for QED and in \cite{Brambilla:2009cd} 
for QCD in the static limit. 
The potential obtained in this way differs in many respects from the 
most commonly used phenomenological potentials (for some reviews see 
\cite{Petreczky:2005bd,Satz:2005hx}). Most remarkably, it develops an imaginary part.
At least two mechanisms have been identified that are responsible for the appearance 
of a thermal width: the Landau-damping phenomenon  \cite{Laine:2006ns}
and the quark-antiquark colour-singlet break up \cite{Brambilla:2009cd}. 
In particular, it has been pointed out that quarkonium dissociation due 
to the former rather than screening may be the dominant 
mechanism at the origin of  heavy quarkonium dissociation in a medium
\cite{Escobedo:2008sy,Laine:2008cf}.
These developments motivate us to revisit
the physics of heavy quarkonium states in a thermal bath in a more systematic way.  
We shall focus here on temperatures for which $\pi T$
is smaller than the typical momentum transfer in the bound states:
such temperatures are those reachable at present days colliders
\cite{Satz:2005hx}.

Heavy quarkonium in a medium is characterized both by the scales
typical of a non-relativi\-stic bound state and by the thermal scales.
The non-relativistic scales are the inverse of the typical radius of
the system $1/a_0$ and its typical binding energy $E$. The thermal
scales are the temperature (or multiple of $\pi T$) and the electric
screening mass $m_D$, among other lower energy scales, which are not
relevant to our discussion.
In the weak-coupling regime, which we will assume throughout this
work, these scales may be expressed in terms of the strong coupling constant
$g\ll 1$, the heavy quark mass $m$, and the temperature $T$: 
$m_D \sim gT$, $1/a_0 \sim m \als$ and $E \sim m \als^2$, where $\als = g^2/(4\pi)$. 
Non-relativistic scales and thermal scales are hierarchically 
ordered. This allows to investigate the quarkonium properties in a medium 
using the same systematic framework provided by non-relativistic effective 
field theories at zero temperature \cite{Brambilla:2004jw}.

In this work, we aim at studying heavy quarkonium at finite temperature 
including the contribution induced by a large but finite quark mass, in this way 
merging and completing the findings of Refs.~\cite{Escobedo:2008sy,Brambilla:2009cd}.
We will adopt the same real-time EFT framework of \cite{Escobedo:2008sy,Brambilla:2009cd} 
and assume for definitiveness the following hierarchy between the thermodynamical 
and the non-relativistic scales:
\begin{equation}
m \gg m\als \gg T \gg m\als^2 \gg m_D.
\label{hierarchy}
\end{equation}
With this choice, the thermal bath affects the Coulombic bound state as a small perturbation, 
yet modifying the Coulomb potential. We remark that this temperature is below the melting temperature, 
which is of order $m\als^{2/3}$ \cite{Escobedo:2008sy}. Moreover, this may indeed 
correspond to the situation of interest in present day colliders.
As a consequence of \eqref{hierarchy}, in the weak-coupling regime, we have
that $mg^3\gg T \gg mg^4$, corresponding to $m g^4\gg m_D\gg mg^5$. 
We furthermore assume that $\Lambda_{\rm QCD}$, the QCD scale, is smaller 
than $m_D$ (although results that do not involve a weak-coupling expansion at the scale 
$m_D$, which are all the results of the paper before Sec.~\ref{secmD}, 
are valid also for $m_D \sim \Lambda_{\rm QCD}$). 
A number of different inequalities has been addressed in the Abelian case in \cite{muonic}. 

We will concentrate on the energy levels and decay widths and we will determine 
how they get modified in a thermal bath whose temperature is such that
it satisfies the conditions \eqref{hierarchy}. In order to be definite, we will 
further assume $(m_D/E)^4 \ll g$, in this way keeping small the number of required  
corrections suppressed by powers of $m_D/E$, and we will evaluate the spectrum with an accuracy 
of order $m\als^5$. 

The strongest limitation for the practical application of our final
results to actual bottomonium and charmonium systems comes from the
fact that we use perturbation theory at the ultrasoft scale
$m\als^2$. Still, we expect them to be relevant for the ground states
of bottomonium and, to a lesser extent, charmonium.  Some intermediate
expressions, for which perturbation theory is only used at the scale
$T\gg m\als^2$ may have a wider range of applicability. We also assume a vanishing charm quark 
mass in the bottomonium case (effects of a non-vanishing mass are discussed in \cite{muonic}).

The paper is organized in the following way. In Sec.~\ref{secrealtime}, we briefly review 
the Feynman rules of QCD at finite temperature in the real-time formalism.
In Sec.~\ref{secm}, we set up the effective field theory that follows from QCD by integrating 
out the scales $m$ and $m\als$ in the heavy quark-antiquark sector. In Sec.~\ref{secT}, 
we calculate the contributions to the spectrum coming from the scale $T$, in Sec.~\ref{secE}, those 
coming from the scale $E$ and, finally, in Sec.~\ref{secmD}, those coming from the scale $m_D$.
In Sec.~\ref{secconclusions}, we summarize our results giving the thermal energy shifts and 
widths up to order $m\als^5$.

\section{QCD at finite temperature in the real-time formalism}
\label{secrealtime}
In this section, we review the Feynman rules of QCD with static quarks at finite temperature in the
real-time formalism.

Real-time expectation values depend on how the contour of the time integration
in the partition function is deformed to include real times.
In the paper, we adopt a contour that goes from an initial time $t_{\rm i}$  to a real final time $t_{\rm f}$, 
from $t_{\rm f}$ to $t_{\rm f}-i0^+$, from  $t_{\rm f}-i0^+$ to $t_{\rm
  i}-i0^+$ and from $t_{\rm i}-i0^+$ to $t_{\rm i}-i/T$.
The propagators will be given with this conventional choice of contour.
Since the contour has two lines moving along the real time axis, degrees of
freedom double in real time and propagators are represented by $2\times 2$ matrices.  
Furthermore we define 
\begin{equation}
n_{\rm B}(k_0) = \frac{1}{e^{k_0/T}-1},
\label{bose}
\end{equation}
which is the Bose--Einstein distribution.

The non-relativistic propagator of an unthermalised quark-antiquark pair interacting through a potential $V(r)$ 
reads (see \cite{Brambilla:2009cd}, we have added here the kinetic energy) 
\begin{equation}
S(k_0,k) =
\left(
\begin{matrix}
&&\hspace{-2mm}\displaystyle\frac{i}{k_0 - k^2/m  - V(r) + i\eta}  
&&0 \\ 
&&2\pi\delta(k_0 - k^2/m - V)  
&&\displaystyle\frac{-i}{k_0  - k^2/m  - V(r) -i\eta}
\end{matrix}
\right).
\label{Squarkstatic}
\end{equation}
The expression for $V(r)$ depends on whether the quark-antiquark pair is in a color singlet 
or in a color octet configuration. 
In the last case, an identity matrix in color space must be understood, and, 
in either case, an identity matrix in spin space is implicit. 
Since the $[S(k_0,k)]_{12}$ component vanishes, 
the quark-antiquark fields of type ``2'' never enter in any amplitude of physical fields,
i.e. fields of type ``1''. As a consequence, the fields ``2'' decouple
and may be ignored when considering physical amplitudes. 

Throughout this paper we adopt the Coulomb gauge for our calculations. The
free gluon propagator reads \cite{Landshoff:1992ne}:
\begin{eqnarray}
 D^{(0)}_{00}( k) &=& 
\left(
\begin{matrix}
&&\hspace{-2mm} \displaystyle \frac{i}{k^2}
&&0
\\
&&\hspace{-2mm} 0
&&\displaystyle -\frac{i}{k^2}
\end{matrix}
\right), 
\label{D000}
\\
\nonumber D^{(0)}_{ij}(k_0,k) &=& 
\left(\delta_{ij} - \frac{k^ik^j}{k^2}\right)
\!\left\{\!
\left(
\begin{matrix}
&&\hspace{-2mm} \displaystyle \frac{i}{k_0^2-k^2 + i\eta}
&&\theta(-k_0) \, 2\pi\delta(k_0^2-k^2)
\\
&&\hspace{-2mm} \theta(k_0) \, 2\pi\delta(k_0^2-k^2)
&&\displaystyle -\frac{i}{k_0^2-k^2 - i\eta}
\end{matrix}
\right)
\right.\\
&&\hspace{3cm}+ 2\pi\delta(k_0^2-k^2)\, n_{\rm B}(|k_0|)\,
\left(
\begin{matrix}
&&\hspace{-2mm} 1
&& 1
\\
&&\hspace{-2mm} 1
&& ~1
\end{matrix}
\right)\!
\Bigg\}\,,
\label{D0ij}
\end{eqnarray}
where  $k$ stands for the modulus of the three momentum $k^i$.

Due to the decoupling of the heavy quarks of type ``2'', in the following we
will need only the ``11'' component of the heavy quark-antiquark propagators. 
At the order we are calculating, this is also the case for the gluon propagators, which will be needed up to one loop. 
All our equations will refer to this component, unless explicitly stated otherwise.
In particular, we recall that at equilibrium the ``11'' component of the gluon propagator  
can be written in terms of the retarded (R) and advanced (A) propagators,
\begin{eqnarray}
&& D_{\mu\nu}^{\rm R}(k) = \int d^4x \, e^{i(k_0x_0-k\cdot x)} \, 
\theta(x_0)\langle [A_\mu(x),A_\nu(0)]\rangle ,
\label{DR}
\\
&& D_{\mu\nu}^{\rm A}(k) = - \int d^4x \, e^{i(k_0x_0-k\cdot x)} \,
\theta(-x_0) \langle [A_\mu(x), A_\nu(0)]\rangle ,
\label{DA}
\end{eqnarray}
as
\begin{equation}
[D_{\mu\nu}]_{11}= \frac{ D_{\mu\nu}^{\rm R}(k_0,k) + D_{\mu\nu}^{\rm A}(k_0,k)}{2}
+ \left(\frac{1}{2} + n_{\rm B}(k_0)\right)\left(D_{\mu\nu}^{\rm R}(k_0,k) - D_{\mu\nu}^{\rm A}(k_0,k)\right),
\label{11component}
\end{equation}
which holds for the tree level propagator as well as for the full one. 
The second term on the right-hand side, proportional to the difference 
between the retarded and advanced propagators, 
is often termed the symmetric propagator.

\section{Integrating out the scales $m$ and $m\als$}
\label{secm}
Our aim is to calculate the quarkonium spectrum in a thermal bath of 
temperature $T$. We take advantage of the hierarchy of scales \eqref{hierarchy} 
by constructing a hierarchy of effective field theories that follow 
from QCD by systematically integrating out the largest scales.
The EFTs are constructed as series of operators whose matrix elements 
scale like the lower scales and that are suppressed by powers of the 
large scales, which have been integrated out.

The first scale to be integrated out from QCD is the heavy quark mass $m$. 
In the matching procedure, smaller scales are expanded. 
Thus, at this stage, the presence of the thermal
scales does not affect the matching of the Lagrangian, which is the 
Lagrangian of non-relativistic QCD (NRQCD) \cite{Caswell:1985ui}.

The next scale to be integrated out is the scale of the inverse of the typical 
distance of the heavy quark and antiquark, which is of order $m\als$.
According to  \eqref{hierarchy}, it is larger than the temperature. 
We are thus allowed to integrate out $m\als$ from NRQCD
setting to zero all thermodynamical scales. Furthermore, under the assumption that
$m\als \gg \Lambda_\mathrm{QCD}$, this integration can be carried out in
perturbation theory order by order in $\als$. The EFT we obtain is
potential non-relativistic QCD (pNRQCD) \cite{Pineda:1997bj,Brambilla:1999xf}.
Its Lagrangian reads
\begin{eqnarray}
{\cal L}_{\textrm{pNRQCD}} &= 
& 
- \frac{1}{4} F^a_{\mu \nu} F^{a\,\mu \nu} 
+ \sum_{i=1}^{n_f}\bar{q}_i\,iD\!\!\!\!/\,q_i 
+ \int d^3r \; {\rm Tr} \,  
\Biggl\{ {\rm S}^\dagger \left[ i\partial_0 - h_s \right] {\rm S} 
+ {\rm O}^\dagger \left[ iD_0 -h_o \right] {\rm O} \Biggr\}
\nonumber\\
&& \hspace{-1.5cm}
+ V_A\, {\rm Tr} \left\{  {\rm O}^\dagger \br \cdot g\be \,{\rm S}
+ {\rm S}^\dagger \br \cdot g\be \,{\rm O} \right\} 
+ \frac{V_B}{2} {\rm Tr} \left\{  {\rm O}^\dagger \br\cdot g\be \, {\rm O} 
+ {\rm O}^\dagger {\rm O} \br \cdot g\be  \right\}  + \dots\,.
\label{pNRQCD}	
\end{eqnarray}
The fields $\mathrm{S}=S\,\mathbf{1}_c/\sqrt{N_c}$ and
$\mathrm{O}=O^a\,T^a/\sqrt{T_F}$, are the quark-antiquark singlet 
and octet fields respectively, $n_f$ is the number of light quarks $q_i$, 
$N_c=3$ is the number of colours, $T_F=1/2$,
$\be$ is the chromoelectric field ($E^i=F^{i0}$)
and $iD_0 \mathrm{O} =i\partial_0 \mathrm{O} - gA_0 \mathrm{O} + \mathrm{O} gA_0$. 
The trace is intended over colour and spin
indices. Gluon fields depend only on the center-of-mass coordinate $\brg$ and on time;
this is achieved by a multipole expansion in the relative
distance $r$. The dots in the last line of Eq. \eqref{pNRQCD} stand
for higher orders in this expansion.

The dependence on the hard and
soft scales $m$ and $1/r$ is encoded in the Wilson coefficients; $V_A$
and $V_B$ are at leading order $V_A=V_B=1$, whereas the singlet and
octet Hamiltonians have the form 
\begin{equation}
h_{s,o}=\frac{\bp^2}{m}
+V^{(0)}_{s,o}
+\frac{V^{(1)}_{s,o}}{m}+\frac{V^{(2)}_{s,o}}{m^2}+\ldots,
\label{sinoctham}
\end{equation}
where $m$ is the heavy quark mass, $\bp=-i\nabla_\br$. 
The dots stand for higher-order terms in the expansion in $1/m$, both for the kinetic
terms (relativistic corrections) and for the potentials, 
as well as for terms that depend on the center of mass three momentum.

The static potentials read
\begin{equation}
V^{(0)}_s=-C_F\frac{\alVs}{r}\,,\qquad V^{(0)}_o=\frac{1}{2N_c}\frac{\alVo}{r}\,,
\label{staticpot}
\end{equation}
where $C_F=(N_c^2-1)/(2N_c)$ and $\alVs$ and $\alVo$ are series in
$\als$ and at the leading order $\alVs=\alVo=\als$. $\alVs$ is known
up to three loops \cite{Smirnov:2009fh,Anzai:2009tm}, whereas $\alVo$
to two loops \cite{Kniehl:2004rk}. Starting from order $\als^4$,
$\alVs$ is infrared divergent. This divergence was first identified in
\cite{Appelquist:1977es} and analyzed in the framework of pNRQCD in
\cite{Brambilla:1999qa}, where it was shown to cancel against an ultraviolet 
(UV) divergence coming from the ultrasoft degrees of freedom (the scale
$E$).

The non-static potentials $V^{(1)}_s$ and $V^{(2)}_s$ can be
read from \cite{Brambilla:1999xj,Brambilla:2004jw}. $V^{(2)}_s$
consists of a sum of many terms, such as a $\bp$-dependent term, terms
depending on the angular momentum, on the heavy quark-antiquark spins
and a spin-orbit term. Some of these terms, as well as $V^{(1)}_s$,
have an infrared divergence. The leading logarithmic dependence on
$\ln (\mu r)$ accompanying these divergences can be read from
\cite{Brambilla:1999xj,Kniehl:2002br}.
In the octet sector, the non-static potentials are not known beyond tree level. 
Fortunately, for the present analysis only the
leading order expression in $\als$ for the static octet potential will
be needed.

The power-counting of the pNRQCD Lagrangian \eqref{pNRQCD} goes as following:
the relative momentum $\bp$ and the inverse distance
$1/r$ have a size of $\order{m\als}$, whereas the time derivative, the 
gluon fields and the center-of-mass momentum $\bpg$ scale like the lower 
energy scales. Therefore, the largest term in the singlet potential 
expanded in $\als$ and $1/m$ is the Coulomb potential $-\cf\als/r$. 
The spectrum of the corresponding singlet Hamiltonian is given by the Coulomb levels
\begin{equation}
E_n=-\frac{mC_F^2\als^2}{4n^2}=-\frac{1}{ma_0^2n^2}\,,\qquad a_0\equiv\frac{2}{m\cf\als}\,.
\label{coulomblevels}
\end{equation}
Subleading terms in the expansions in $\als$ and $1/m$ are 
treated in quantum-mechanical perturbation theory. The corresponding
shifts of the Coulomb levels have been computed in
\cite{Brambilla:1999xj,Kniehl:2002br,Penin:2002zv}. The infrared
divergences mentioned above affect the spectrum at order $m\als^5$.

For what concerns the propagators of the singlet and octet fields in
the real-time formalism, we have shown in Eq. \eqref{Squarkstatic}
that the "12" component of a quark-antiquark propagator in a potential
$V(r)$ vanishes and the unphysical "2" component decouples. We are
thus allowed to drop also here the real-time formalism indices and
write only the "11" component of the propagator. For the rest of the
paper, all amplitudes will be intended as the "11" components of the
real-time matrices unless otherwise specified.
In particular, for what concerns the singlet propagator, we thus have
\begin{equation}
S^{\rm singlet}(E) =\frac{i}{E -h_s+i\eta},
\label{singletprop}
\end{equation}
where $E$ is the singlet energy. From Eq. \eqref{sinoctham}, the singlet Hamiltonian $h_s$ reads 
\begin{equation}
h_s=\frac{\bp^2}{m}-\cf\frac{\als}{r} + \cdots \;,
\label{leadingham}
\end{equation}
where the dots stand for higher-order terms.
In order to have a homogeneous power counting in the propagator, 
it is convenient to expand it around the leading-order Hamiltonian
\eqref{leadingham}, which is of order  $m\als^2$. 
Similarly the octet propagator is
\begin{equation}
S^{\rm octet}(E)_{ab} = \int_0^\infty dt \;  e^{i(E-h_o)t}\left( e^{-ig\int_{0}^{t}A_0}\right)_{ab},
\label{octetprop}
\end{equation}
where the octet Hamiltonian $h_o$ reads 
\begin{equation}
h_o=\frac{\bp^2}{m}+\frac{1}{2\nc}\frac{\als}{r} + \cdots\;,
\label{leadinghamoctet}
\end{equation}
and the dots stand for terms smaller than $m\als^2$. 
The Wilson line in (\ref{octetprop}) can be expanded in powers of $g$.
We will only need the leading order in such expansion,
\begin{equation}
S^{\rm octet}(E)_{ab} =\frac{i\delta_{ab}}{E -h_o+i\eta}.
\label{octetprop2}
\end{equation}

\section{Integrating out the temperature}
\label{secT}
In this section, we proceed to integrate out modes of energy and
momentum of the order of the temperature $T$. This amounts to
modifying pNRQCD into a new EFT where only modes with energies
and momenta lower than $T$ are dynamical. We may denote the new EFT with 
$\textrm{pNRQCD}_\mathrm{HTL}$ \cite{Vairo:2009ih}. The EFT can be used for $m\als\gg T\gg E,m_D$ 
no matter what the relation between $E$ and $m_D$ is. Its Lagrangian will get additional
contributions with respect to pNRQCD. For our purposes, we are
interested in the modifications to the singlet sector, corresponding
to a thermal correction $\delta V_s$ to the singlet potential, and to the Yang--Mills
sector, amounting to the Hard Thermal Loop (HTL) Lagrangian ${\cal L}_{\rm HTL}$ \cite{Braaten:1991gm}:
\begin{eqnarray}
{\cal L}_{\textrm{pNRQCD}_\mathrm{HTL}} &=& 
{\cal L}_{\rm HTL}
+ \int d^3r \; {\rm Tr} \,  
\Biggl\{ {\rm S}^\dagger \left[ i\partial_0 - h_s - \delta V_s \right] {\rm S} 
+ {\rm O}^\dagger \left[ iD_0 -h_o - \delta V_o \right] {\rm O} \Biggr\}
\nonumber\\
&& \hspace{-1.5cm}
+ {\rm Tr} \left\{  {\rm O}^\dagger \br \cdot g\be \,{\rm S}
+ {\rm S}^\dagger \br \cdot g\be \,{\rm O} \right\} 
+ \frac{1}{2} {\rm Tr} \left\{  {\rm O}^\dagger \br\cdot g\be \, {\rm O} 
+ {\rm O}^\dagger {\rm O} \br \cdot g\be  \right\}  + \dots\,,
\label{pNRQCDHTL}	
\end{eqnarray}
where we have set to one the matching coefficients of the dipole terms, whose 
quantum corrections are beyond the accuracy of the present paper.

\FIGURE[ht]{
\parbox{15cm}{
\centering
\epsfig{file=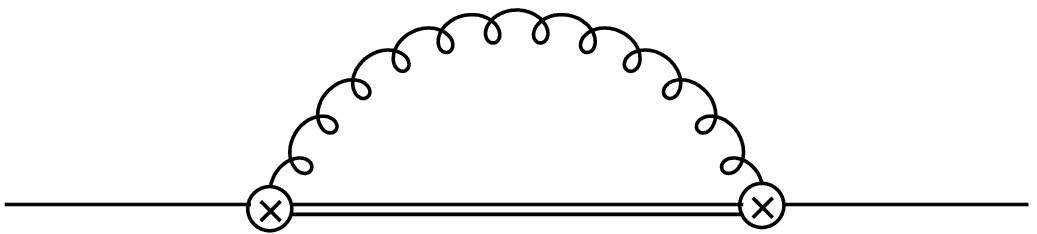,width=7.5cm}
\caption{The leading heavy quarkonium self-energy diagram. 
The single line is a singlet propagator, the double line an octet propagator, 
the curly line a gluon and the vertices are chromoelectric dipoles vertices.}
\label{fig:leading}}
}

We calculate the correction $\delta V_s$ to the singlet potential. 
As in \cite{Brambilla:2009cd,Escobedo:2008sy}, the leading thermal correction is due to the dipole
vertices ${\rm O}^\dagger \br \cdot g\be \,{\rm S} + {\rm S}^\dagger
\br \cdot g\be \,{\rm O}$ in the pNRQCD Lagrangian
\eqref{pNRQCD}. These terms induce the diagram depicted in
Fig.~\ref{fig:leading}, where a colour-singlet state emits and
reabsorbs a chromoelectric gluon through the dipole vertex and an
intermediate colour-octet state. The amplitude reads (see
\cite{Brambilla:1999xj,Kniehl:1999ud,Kniehl:2002br} for the $T=0$
case)
\begin{equation}
- i g^2 \, C_F \, \frac{r^i}{D-1}
\mu^{4-D} \int \frac{d^Dk}{(2\pi)^D}
\frac{i}{E-h_o-k_0 +i\eta}\left[k_0^2 \,  D^{(0)}_{ii}(k_0,k) +  k^2 \,  D^{(0)}_{00}(k_0,k)
\right]r^i\,,
\label{defleading}
\end{equation}
where $E$ is the energy of the singlet; we recall that 
this expression corresponds to the ``11'' component in the real-time formalism. 
Integrals over momenta are regularized in dimensional
regularization, with $D=4+\epsilon$ and $\mu$ being the subtraction point. 
In Coulomb gauge, with the free propagators given in
Eqs. \eqref{D000} and \eqref{D0ij}, the contribution of the
longitudinal gluon vanishes in dimensional regularization, whereas
that of the transverse gluon can be divided into a vacuum and a
thermal part:
\begin{eqnarray}
\nonumber&& -ig^2C_F\frac{D-2}{D-1} r^i\mu^{4-D}
\int\frac{\,d^Dk}{(2\pi)^D}\frac{i}{E-h_o-k_0+i\eta}k_0^2\left[\frac{i}{k_0^2-k^2+i\eta}\right.\\
&&\hspace{7cm}+2\pi \delta\left(k_0^2-k^2\right)n_\mathrm{B}\left(\vert k_0\vert\right)\bigg]r^i\,;
\label{transverseleading}
\end{eqnarray}
the first term in the square brackets is the vacuum part and the second term is the thermal part. 
The expression depends on the scales $T$ and $E$. In order to single out
the contribution from the scale $T$, which comes from the momentum regions $k_0\sim T$ 
and $k\sim T$, we recall that $T\gg(E-h_o)$ and expand the octet propagator as
\begin{equation}
\frac{i}{E-h_o-k_0+i\eta}=
\frac{i}{-k_0+i\eta}-i\frac{E-h_o}{(-k_0+i\eta)^2}
+i\frac{(E-h_o)^2}{(-k_0+i\eta)^3}-i\frac{(E-h_o)^3}{(-k_0+i\eta)^4}+\ldots\,.
\label{octetexpand}
\end{equation}
The contribution of the vacuum part of the propagator is
scaleless for all the terms of the expansion and thus it vanishes. 
Conversely, in the thermal part, we have the Bose--Einstein
distribution giving a scale to the integration.

The zeroth-order term in the expansion \eqref{octetexpand} gives a vanishing
integral \cite{Brambilla:2009cd}, whereas the following terms contribute to the
potential. The linear and the cubic terms in $E-h_o$, i.e. 
\begin{eqnarray}
&&-g^2C_F\frac{D-2}{D-1}r^i(E-h_o)r^i\,\mu^{4-D}\int\frac{\,d^{D-1}k}{(2\pi)^{D-1}}\frac{n_\mathrm{B}(k)}{k}, 
\label{deflinear}
\end{eqnarray}
and 
\begin{eqnarray}
&&-g^2C_F\frac{D-2}{D-1}r^i(E-h_o)^3r^i\,\mu^{4-D}\int\frac{\,d^{D-1}k}{(2\pi)^{D-1}}\frac{n_\mathrm{B}(k)}{k^3},
\label{defcubic}
\end{eqnarray}
can be shown to contribute to the real part of the potential.
Since in our counting \eqref{deflinear} behaves as $mg^{8} \gg \als T^2 Er^2\gg mg^{10}$ 
and \eqref{defcubic} as $\als E^3r^2\sim mg^{10}$, further terms
in the ${E}/{T}$ expansion are not needed. 
Finally, the square term in the expansion, which would give an imaginary contribution to the potential, 
vanishes in dimensional regularization:
\begin{equation}
\frac{g^2C_F}{2}\frac{D-2}{D-1}r^i(E-h_o)^2r^i\,\mu^{4-D}
\int\frac{\,d^{D-1}k}{(2\pi)^{D-1}}\,k\,n_\mathrm{B}(k)\left[\frac{1}{(-k+i\eta)^3}+\frac{1}{(k+i\eta)^3}\right]
=0.
\label{defimag}
\end{equation}

We now evaluate the linear term defined in Eq. \eqref{deflinear}. 
The integration yields
\begin{equation}
\label{linearwithenergy}
-\frac{2\pi}{9}C_F\als T^2\,r^i(E-h_o)r^i\,.
\end{equation}
Matching the singlet propagator in pNRQCD with the singlet propagator in $\textrm{pNRQCD}_\mathrm{HTL}$ 
we obtain 
\begin{equation}
\frac{1}{E-h_s} + \frac{1}{E-h_s}\left(-\frac{2\pi}{9}C_F\als T^2\,r^i(E-h_o)r^i\right)\frac{1}{E-h_s}
=\delta Z_s^{1/2}\frac{1}{E-h_s-\delta V_s}\delta Z_s^{1/2\,\dagger}\,,
\label{matchcond}
\end{equation}
where the left-hand part of the equality corresponds to the pNRQCD part of the 
matching and the right-hand side to the  $\textrm{pNRQCD}_\mathrm{HTL}$ part of the matching:
$\delta V_s$ is the thermal correction to the singlet potential in $\textrm{pNRQCD}_\mathrm{HTL}$
and $\delta Z_s$ the thermal correction to the singlet normalization in $\textrm{pNRQCD}_\mathrm{HTL}$. 
Our purpose is solely the evaluation of $\delta V_s$, which is necessary for the
spectrum. So we rewrite $E-h_o$ as $E-h_s-(h_o-h_s)$, where
$(h_o-h_s)$ is given by the difference between the octet and singlet potentials:
\begin{equation} 
h_o-h_s=\sum_n\frac{V_o^{(n)}-V_s^{(n)}}{m^n}\equiv\Delta V\,.
\label{deltaV}
\end{equation}
$\Delta V$ is organized as an expansion in $\als$ and $1/m$. 
At the leading order, it is the difference between the tree-level static potentials:
\begin{equation}
\Delta V=\frac{\nc}{2}\frac{\als}{ r}.
\label{leadingdv}
\end{equation}
Higher-order terms are easily shown to contribute to the spectrum beyond our accuracy. 
Similarly, for what concerns the singlet Hamiltonian, only the leading terms displayed in
Eq. \eqref{leadingham} are necessary. Hence $r^i(E-h_o)r^i$ simplifies
into $r^i(E-h_s)r^i- \nc\als r/2$; the second term is easily identified as contributing 
to $\delta V_s$, whereas plugging the first term back into Eq. \eqref{matchcond} yields
\begin{equation}
\label{commutexample}
\frac{1}{E-h_s}r^i(E-h_s)r^i\frac{1}{E-h_s}=
\frac{1}{2}\frac{1}{E-h_s}\left(\left[[r^i,E-h_s],r^i\right]+\{r^2,(E-h_s)\}\right)\frac{1}{E-h_s}. 
\end{equation}
The term $\{r^2,(E-h_s)\}$ contributes to the normalization of the wave function. We are thus left with
\begin{equation}
\label{commutexample2}
\frac{1}{E-h_s}\frac{1}{2}[[h_s,r^i],r^i]\frac{1}{E-h_s}.
\end{equation}
The commutator can be easily computed from the Hamiltonian displayed in Eq. \eqref{leadingham}. 
Plugging the result back into Eq. \eqref{matchcond}, we obtain from the matching condition
\begin{equation}
\delta V_s^{\rm (linear)}=\frac{\pi}{9}N_c C_F \als^2 T^2r+\frac{2\pi}{3m}C_F\als T^2\,.
\label{deltavslinear}
\end{equation}
The first term is the contribution of $\Delta V$ and was first obtained in \cite{Brambilla:2009cd}. 
The second term is the contribution of the kinetic term; a similar term appears in the
Abelian case of Ref.~\cite{Escobedo:2008sy}.
Using first-order quantum-mechanical perturbation theory and the
expectation values $\left\langle r \right\rangle_{n,l}$ on the eigenstates of the
Coulomb potential ($n$ and $l$ stand for the principal 
and angular momentum quantum numbers respectively, see, for instance, \cite{Titard:1993nn}) 
we obtain the following correction to the Coulomb energy levels
\begin{equation}
\delta E_{n,l}^{\rm (linear)}=\frac{\pi}{9}N_c C_F \als^2 T^2 \frac{a_0}{2}[3n^2-l(l+1)]+\frac{2\pi}{3m}C_F\als T^2.
\label{linear}
\end{equation}

We now move to the cubic term, as defined in Eq. \eqref{defcubic}. We have
\begin{equation}
-g^2C_F\frac{D-2}{D-1}r^i(E-h_o)^3r^i\,\mu^{4-D}
\int\frac{\,d^{D-1}k}{(2\pi)^{D-1}}\frac{n_\mathrm{B}(k)}{k^3}=\frac{\als C_FI_T}{3\pi}r^i(E-h_o)^3r^i\,,
\label{cubic}
\end{equation}
where $I_T$ comes from the evaluation of the integral. It reads \cite{Escobedo:2008sy} 
\begin{equation}
I_T=\frac{2}{\epsilon}+\ln\frac{T^2}{\mu^2}-\gamma_E+\ln( 4\pi)-\frac{5}{3}\,,
\end{equation}
where $\gamma_E$ is the Euler's gamma.
The divergence of this expression is of infrared (IR) origin: it arises
when integrating over the Bose--Einstein distribution at momenta much
smaller than the temperature. Since we are integrating out the
temperature, i.e. getting the contribution for $k\sim T$, this
divergence is an artifact of our scale separation. We identify two
possible schemes in which the cancellation of this divergence may be interpreted.
\begin{enumerate}
        \item In the first scheme, the divergence is cancelled by an
opposite ultraviolet divergence from a lower scale, in our case the
binding energy. In the next section, we will indeed show that
the thermal part of this very same diagram, when evaluated for
loop momenta of the order of the binding energy, yields 
an ultraviolet divergence that exactly cancels the one here, 
whereas the vacuum part of that diagram gives an
opposite UV divergence that cancels the IR divergence of the pNRQCD
potentials, yielding a finite spectrum.
	\item Alternatively one can observe that the pole of the
divergence is exactly opposite to the infrared pole of the pNRQCD potentials,
which can be read from \cite{Brambilla:1999xj} and the two therefore
cancel. More precisely, the scaleless, and hence vanishing in dimensional regularization, integral of 
the vacuum part of Eq. \eqref{transverseleading}, with the octet propagator 
expanded at the cubic order, can be rewritten  
as the sum of an infrared and an ultraviolet divergent integral. 
The infrared pole cancels with the one in Eq. \eqref{cubic} coming from the thermal
part, whereas the ultraviolet one cancels the IR divergence of the pNRQCD 
potentials. 
\end{enumerate}
The two interpretation schemes are equivalent and produce at the end a finite
spectrum, which is the relevant observable. 

The evaluation of $r^i(E-h_o)^3r^i$ in (\ref{cubic}), in analogy to what has been performed previously in
Eqs. \eqref{commutexample} and \eqref{commutexample2}, can be read from \cite{Kniehl:2002br}
\begin{eqnarray}
\nonumber &&\frac{1}{E-h_s}r^i(E-h_o)^3r^i\frac{1}{E-h_s}=
\frac{1}{E-h_s}\left(-\frac{N_c^3}{8}\frac{\als^3}{r}-(\nc^2+2\nc\cf)\frac{\als^2}{mr^2}\right.\\
\nonumber&&\left.+4(\nc-2\cf)\frac{\pi\als}{m^2}\delta^3(\br)+\nc\frac{\als}{m^2}
\left\{\nabla^2_\br,\frac{1}{r}\right\}\right)\frac{1}{E-h_s}+\cdots ,\\
\label{cubicexp}
\end{eqnarray}
where the dots stand for wave function renormalizations.
Matching to the right-hand side of Eq. \eqref{matchcond}, we obtain 
the corresponding contribution to the singlet potential 
$\delta V_s$ of $\textrm{pNRQCD}_\mathrm{HTL}$:
\begin{eqnarray}
\nonumber\delta V_s^{\rm (cubic)}&=&
\frac{\als C_FI_T}{3\pi}\left(-\frac{N_c^3}{8}\frac{\als^3}{r}-(\nc^2+2\nc\cf)\frac{\als^2}{mr^2}\right.\\
&& \hspace{2cm}
\left.+4(\nc-2\cf)\frac{\pi\als}{m^2}\delta^3(\br)+\nc\frac{\als}{m^2}\left\{\nabla^2_\br,\frac{1}{r}\right\}\right)\,.
\label{cubicpot}
\end{eqnarray}
Using first-order quantum-mechanical perturbation theory 
and the value of the Coulomb wave function at the origin, 
$\left\vert\psi_{n,l}(0)\right\vert^2=\delta_{l0}/(\pi n^3a_0^3)$, 
we obtain the shift of the energy levels
\begin{equation}
\delta E_{n,l}^{\rm (cubic)}=
\frac{E_nI_T\als^3}{3\pi}\left\{\frac{4 C_F^3\delta_{l0} }{n}+N_c \cf^2\left(\frac{8}{n (2l+1)}-\frac{1}{n^2} 
- \frac{2\delta_{l0}}{ n }   \right)+\frac{2N_c^2C_F}{n(2l+1)}+\frac{N_c^3}{4}\right\}\,.
\label{cubicshift}
\end{equation}

\FIGURE[ht]{
\parbox{15cm}{
\centering
\epsfig{file=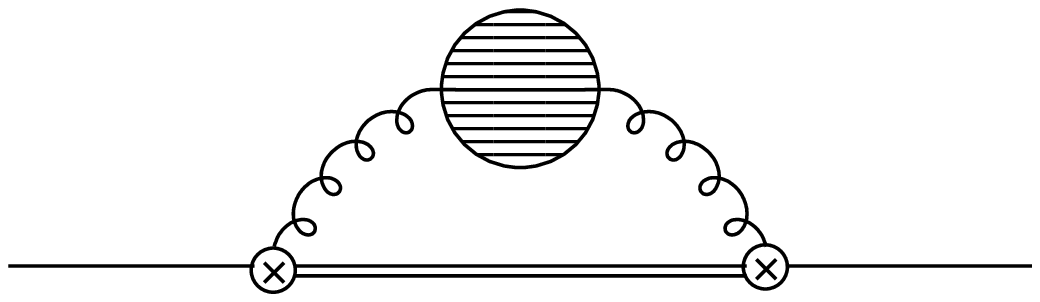,width=8cm}
\caption{Contribution from the NLO chromoelectric correlator: 
the gluon is longitudinal and the dashed blob is the one-loop self-energy.}
\label{fig:loop}}
}

Another possible contribution to the potential up to order 
$m\als^5$ is given by radiative corrections to the diagram
shown in Fig.~\ref{fig:leading}. At the next order in $\als$,
corresponding to two loops, a sizable number of diagrams appears. In
\cite{Brambilla:2009cd}, it was shown that in Coulomb gauge only one diagram needs to be
considered. It consists of a one-loop self-energy insertion in the
longitudinal part of the chromoelectric correlator and it is shown in
Fig.~\ref{fig:loop}.  It contributes at order $\als Tm_D^2 r^2$,
corresponding, in our scale hierarchy, to a magnitude in between $mg^9$ and 
$mg^{12}$. Therefore, this term contributes to the spectrum up to order 
$m\als^5$ only if $mg^3\gg T\ge mg^{10/3}$. 
This makes clear that non-static contributions that were
not considered in the analysis of Ref.~\cite{Brambilla:2009cd}, such as 
vertices originating from the spatial center-of-mass covariant
derivative in the octet sector and higher-order singlet-octet vertices 
in the $1/m$ expansion (see \cite{Brambilla:2003nt}), contribute to terms smaller than $m\als^5$ only.
At our accuracy, the diagram we are considering can again be written expanding  
the octet propagator for $k_0\sim T\gg(E-h_o)$ and retaining only the first term, independent 
of $E-h_o$. Therefore, the result is the same as the one derived in \cite{Brambilla:2009cd}. It reads
\begin{eqnarray}
\delta  V_s^{\rm (2 \, loops)}
&=& 
 - \frac{3}{2} \zeta(3)\,  C_F \, \frac{\als}{\pi} \, r^2 \, T \,m_D^2
+ \frac{2}{3} \zeta(3)\, N_c C_F \, \als^2 \, r^2 \, T^3 
\nonumber\\
&&
+ i \left[ \frac{C_F}{6} \als \, r^2 \, T \,m_D^2\, \left( 
-\frac{2}{\epsilon} + \gamma_E + \ln\pi 
- \ln\frac{T^2}{\mu^2} + \frac{2}{3} - 4 \ln 2 - 2 \frac{\zeta^\prime(2)}{\zeta(2)} \right)\right.
\nonumber\\
&&  \quad \left.
+ \frac{4\pi}{9} \ln 2 \; N_c C_F \,  \als^2\, r^2 \, T^3 \right]
\,,
\label{VsTloop}
\end{eqnarray}
where $\zeta$ is the Riemann zeta function ($\zeta(2) = \pi^2/6$) and the Debye mass $m_D$
is defined as
\begin{equation}
m_D^2 = \frac{g^2T^2}{3}\left(N_c + T_F\,n_f \right).
\label{mD}
\end{equation}

Equation \eqref{VsTloop} contains an imaginary part. It comes from the
imaginary part of the gluon self-energy, which is related to the 
Landau-damping phenomenon, i.e. the scattering of particles carrying momenta of order $T$ 
in the thermal bath with virtual, space-like
longitudinal gluons. Furthermore, the imaginary part is infrared
divergent. In the EFT framework, this divergence has to be cancelled by an opposite
ultraviolet divergence coming from a lower scale. In the following section, we
will indeed show that the same diagram, when integrated over momenta
of the order of the binding energy, yields the desired UV divergence. 
Finally, we remark that the result in Eq. \eqref{VsTloop} comes from 
dimensionally regularizing only the integral over $k$ while keeping 
the thermal part of the gluon self energy, which is finite, 
in exactly four spacetime dimensions. Using the same regularization when 
calculating the contribution coming from the binding-energy scale guarantees 
that the final result for the spectrum is finite and scheme independent.
This is not the case for the potential, however, whose expression 
depends on the adopted scheme.

The contributions to the energy levels and to the thermal width can be
obtained easily from Eq. \eqref{VsTloop} by using the expectation
value for $r^2$ on Coulombic states, i.e. $\left\langle
r^2\right\rangle_{n,l}= a_0^2n^2\left[5n^2+1-3l(l+1)\right]/2$:
\begin{eqnarray}
\hspace{-0.7cm} 
\delta  E_{n,l}^{\rm (2 \, loops)} &=& 
\left[ - \frac{3}{4} \zeta(3)\,  C_F \, \frac{\als}{\pi} \, T \,m_D^2
+ \frac{\zeta(3)}{3}  N_c C_F \, \als^2 \, T^3 \right]
a_0^2n^2\left[5n^2+1-3l(l+1)\right],
\label{EnTloop}
\\
\hspace{-0.7cm} 
\Gamma_{n,l}^{\rm (2 \, loops)} &=& 
\left[ - \frac{C_F}{6} \als T m_D^2
\left( -\frac{2}{\epsilon} + \gamma_E + \ln\pi 
- \ln\frac{T^2}{\mu^2}+ \frac{2}{3} - 4 \ln 2 - 2 \frac{\zeta^\prime(2)}{\zeta(2)} \right) \right.
\nonumber\\
&& \left.
-\frac{4\pi}{9} \ln 2 \; N_c C_F \,  \als^2\, T^3 \right] {a_0^2n^2}\left[5n^2+1-3l(l+1)\right]\,.
\label{GammanTloop}
\end{eqnarray}

\subsection{Summary}
Summing up Eqs. \eqref{deltavslinear}, \eqref{cubicpot} and \eqref{VsTloop} 
we obtain the thermal correction to the potential in $\textrm{pNRQCD}_\mathrm{HTL}$
up to terms whose contribution to the spectrum is smaller than $m\als^5$: 
\begin{eqnarray}
\nonumber
\delta V_s &=&\frac{\pi}{9}N_c C_F\, \als^2\, T^2\,r+\frac{2\pi}{3m}C_F\,\als \,T^2
+\frac{\als C_FI_T}{3\pi}\left[-\frac{N_c^3}{8}\frac{\als^3}{r}-(\nc^2+2\nc\cf)\frac{\als^2}{mr^2}\right.
\\
\nonumber&&
\hspace{5cm}
\left.+4(\nc-2\cf)\frac{\pi\als}{m^2}\delta^3(\br)+\nc\frac{\als}{m^2}\left\{\nabla^2_\br,\frac{1}{r}\right\}\right]
\\
&&- \frac{3}{2} \zeta(3)\,  C_F \, \frac{\als}{\pi} \, r^2 \, T \,m_D^2
+ \frac{2}{3} \zeta(3)\, N_c C_F \, \als^2 \, r^2 \, T^3
\nonumber\\
&&
+ i \left[ \frac{C_F}{6} \als \, r^2 \, T \,m_D^2\, \left( 
-\frac{2}{\epsilon} + \gamma_E + \ln\pi 
- \ln\frac{T^2}{\mu^2} + \frac{2}{3} - 4 \ln 2 - 2 \frac{\zeta^\prime(2)}{\zeta(2)} \right)\right.
\nonumber\\
&&  \quad \left.
+ \frac{4\pi}{9} \ln 2 \; N_c C_F \,  \als^2\, r^2 \, T^3 \right]
\,,
\label{totalpotT}
\end{eqnarray}
where the first two terms come from the linear part of
Fig.~\ref{fig:leading}, the terms in square brackets come from the
cubic term and the last three lines originate from the diagram in Fig.~\ref{fig:loop}. 
This correction to the potential can be used for $T\gg E,m_D$ 
no matter what the relative size between $E$ and $m_D$ is.

Analogously, the total contribution to the energy levels coming from the scale $T$ is
\begin{eqnarray}
\nonumber \delta E_{n,l}^{(T)}&=&
\frac{\pi}{9}N_c C_F \,\als^2 \,T^2 \frac{a_0}{2}(3n^2-l(l+1))+\frac{2\pi}{3m}C_F\, \als\, T^2
\\
\nonumber&&
+\frac{E_nI_T\als^3}{3\pi}\left\{\frac{4 C_F^3\delta_{l0} }{n}
+N_c \cf^2\left(\frac{8}{n (2l+1)}-\frac{1}{n^2} - \frac{2\delta_{l0}}{ n }   \right)
+\frac{2N_c^2C_F}{n(2l+1)}+\frac{N_c^3}{4}\right\}
\\
\nonumber&& 
+\left(- \frac{3}{2} \zeta(3)\,  C_F \, \frac{\als}{\pi}  \, T \,m_D^2
+ \frac{2}{3} \zeta(3)\, N_c C_F \, \als^2 \, T^3\right) \frac{a_0^2n^2}{2}\left[5n^2+1-3l(l+1)\right].
\\
\label{totalenergyT}
\end{eqnarray}
The first and the second lines originate from the diagram in Fig.~\ref{fig:leading}, 
and correspond to the linear and cubic terms in the expansion (\ref{octetexpand}). 
The last line originates from the gluon self-energy diagram in Fig.~\ref{fig:loop}, 
which also gives the full contribution of the scale $T$ to the width:
\begin{eqnarray}
\Gamma_{n,l}^{(T)}&=& \Gamma_{n,l}^{\rm (2 \, loops)}.
\label{totalwidthT}
\end{eqnarray}

\section{Contribution to the spectrum from the scale  $E$  \label{sec:energy}}
\label{secE}
After having integrated out the temperature in the previous section,
many different scales ($E$, $m_D$, $\Lambda_{\rm QCD}$, $\ldots$)
still remain dynamical in $\mathrm{pNRQCD}_\mathrm{HTL}$. 
In our hierarchy, the binding energy is much
larger than the Debye mass and $\Lambda_{\rm QCD}$ is smaller than all
other scales. Our purpose is to compute the correction to the
spectrum and the width coming from the scales $E$ and $m_D$. This is
achieved by computing loop corrections to the singlet propagator in 
$\mathrm{pNRQCD}_\mathrm{HTL}$. We recall that the gauge sector of $\mathrm{pNRQCD}_\mathrm{HTL}$ 
coincides with the Hard Thermal Loop effective Lagrangian.
The longitudinal and transverse gluon propagators in Coulomb gauge 
are given in the Hard Thermal Loop effective theory by \cite{Carrington:1997sq}\footnote{The transverse 
propagator given there contains a misprint: a factor of $p_0/(2p)$ should be multiplying 
the logarithm in Eq. (27), as follows from the transverse HTL self-energy given in Eq. (17) 
of the same paper.}
\begin{equation}
D^{\mathrm{R,A}}_{00}(k_0,k)=
\frac{i}{k^2+m_D^2\left(1-\displaystyle\frac{k_0}{2k}\ln\frac{k_0+k\pm i\eta}{k_0-k\pm i\eta}\right)}\,,
\label{prophtllong}
\end{equation}
and 
\begin{equation}
D^{\mathrm{R,A}}_{ij}(k_0,k)=\left(\delta_{ij}-\frac{k_ik_j}{k^2}\right)\Delta_\mathrm{R,A}(k_0,k)\,,
\label{prophtltrans}
\end{equation}
respectively,  where
\begin{equation}
\Delta_\mathrm{R,A}(k_0,k)=
\frac{i}{k_0^2-k^2-\displaystyle\frac{m_D^2}{2}
\left(\displaystyle\frac{k_0^2}{k^2}-(k_0^2-k^2)\displaystyle\frac{k_0}{2k^3}
\ln\left(\displaystyle\frac{k_0+k\pm i\eta}{k_0-k\pm i\eta}\right)\right)\pm i\,\mathrm{sgn}(k_0)\,\eta}\,,
\label{defdelta}
\end{equation}
and the upper sign refers to the retarded propagator and the lower sign to the advanced one. 
The ``11'' component can be obtained from the relation \eqref{11component}.

We start by evaluating the diagram shown in
Fig.~\ref{fig:leading}, whose general expression is given in
Eq. \eqref{defleading}, but now the longitudinal and transverse gluon propagators 
are given by Eqs. \eqref{prophtllong} and \eqref{prophtltrans}.
As we shall see, this is the only diagram we
need to consider to get the spectrum at order $m\als^5$.

At the energy scale, we have $k_0\sim(E-h_o)$ and therefore we have to keep 
the octet propagator unexpanded. However two expansions are still possible.
\begin{enumerate}
        \item Since $k\sim E\ll T$, the Bose--Einstein distribution can be expanded in
\begin{equation}
\label{boseexp}
\frac{1}{e^{k/T}-1}=\frac{T}{k}-\frac{1}{2}+\frac{k}{12 \, T}+\ldots\,.
\end{equation}
        \item Moreover, since  $k\sim E\gg m_D$, the Hard Thermal Loop propagators
can be expanded in $m_D^2/E^2\ll1$. At the zeroth order, this
corresponds to using the propagators given in Eqs. \eqref{D000} and \eqref{D0ij}. 
Some care is required in the expansion of the transverse gluons 
due to a collinear region, as we shall see later on.
\end{enumerate}
In the following, we will call $\delta \Sigma_s(E)$ the contribution 
of the diagram in Fig.~\ref{fig:leading} to the singlet self energy; 
the corresponding energy shift and width for the state $\vert
n,l\rangle$ are given by 
$\delta E_{n,l}=\langle n,l\vert \mathrm{Re}\,\delta \Sigma_s(E_{n,l})\vert n,l\rangle$ and 
$\Gamma_{n,l}=-2\langle n,l\vert \mathrm{Im}\,\delta \Sigma_s(E_{n,l})\vert n,l\rangle$.

We now proceed to the evaluation of Eq. \eqref{defleading} for loop
momenta of the order of the binding energy, with the HTL propagators
defined in Eqs. \eqref{prophtllong} and \eqref{prophtltrans}. We
find convenient to compute separately the contributions coming from the 
transverse and longitudinal gluons.

\subsection{Transverse gluon contribution}
The contribution of transverse gluons to Eq. \eqref{defleading} is in $\mathrm{pNRQCD}_\mathrm{HTL}$
\begin{eqnarray}
\nonumber\delta \Sigma_s^{\rm (trans)}(E) &=&	- i g^2 \, C_F \, \frac{D-2}{D-1}r^i
\mu^{4-D} \!\! \int \!\! \frac{d^Dk}{(2\pi)^D}
\frac{i}{E-h_o-k_0 +i\eta}k_0^2 \left[ \frac{ \Delta_{\rm R}(k_0,k) + \Delta_{\rm A}(k_0,k)}{2}\right.\\
&&\hspace{3.2cm}\left.+ \left(\frac{1}{2} + n_{\rm B}(k_0)\right)\left(\Delta_{\rm R}(k_0,k) 
- \Delta_{\rm A}(k_0,k)\right)\right]r^i\,.
\label{deftransE}
\end{eqnarray}
We start by evaluating the contribution of the symmetric part, which turns out to be the leading one: 
\begin{eqnarray}
\nonumber&& g^2 \, C_F \, \frac{D-2}{D-1}r^i
\mu^{4-D} \int \frac{d^Dk}{(2\pi)^D}
\frac{k_0^2}{E-h_o-k_0 +i\eta} \left(\frac{T}{k_0}+\order{\frac ET}\right)
\\
&&\hspace{7.5cm}
\times\left[\Delta_{\rm R}(k_0,k) - \Delta_{\rm A}(k_0,k)\right]r^i\,,
\label{deftranssym}
\end{eqnarray}
where we have expanded the Bose--Einstein distribution. 
The expansion of the HTL propagators for $m_D\ll k_0,k$
needs to be performed with care in the region around the light cone,
where the gluon propagator becomes singular. We refer to Appendix
\ref{app_trans} for details on the expansion and the evaluation of the
integral, whose final result reads
\begin{equation}
-i\frac{2}{3}\als\,\cf T r^i(E-h_o)^2r^i
+i\frac{\als\cf\,Tm_D^2\,r^2\,(\ln2-1/2)}{3}+\order{\als T m_D^4r^2/E^2,\als r^2E^4/T}\,.
\label{finaltranssym}
\end{equation}
The suppressed term of order $\als r^2 E^4/T$ comes from the $k/(12\,T)$
term in the expansion of the thermal distribution, whereas the term of
order $\als T m_D^4r^2/E^2$ comes from subleading terms in the
expansion of the propagator.\footnote{This term is of order
$m\als^5$ or bigger only in the very tiny window $mg^3\gg T\ge
mg^{3+1/5}$. For this reason, we will not include terms of order $\als T m_D^4r^2/E^2$
or smaller obtained from the expansion in $m_D^2/E^2$. \label{footpre}}

We now consider the first term in the square brackets in
Eq. \eqref{deftransE}; it does not depend on the Bose--Einstein
distribution and, when expanded for $k_0,k\sim E\gg m_D$, gives
\begin{equation}
\frac{ \Delta_{\rm R}(k_0,k) + \Delta_{\rm A}(k_0,k)}{2}=i\mathrm{P}\frac{1}{k_0^2-k^2}+\order{m_D^2/E^4},
\label{r+atrans}
\end{equation}
where P stands for the principal value prescription.
Plugging Eq. \eqref{r+atrans} back into Eq. \eqref{deftransE} yields
\begin{eqnarray}
\nonumber&&	 g^2 \, C_F \, \frac{D-2}{D-1}r^i
\mu^{4-D} \int \frac{d^Dk}{(2\pi)^D}
\frac{1}{E-h_o-k_0 +i\eta}k_0^2 \left[ i\mathrm{P}\frac{1}{k_0^2-k^2}+\order{m_D^2/E^4}\right]r^i\\
&&=-i\frac{\als \, C_F \, }{3}r^i(E-h_o)^3r^i+\order{\als \,E \,m_D^2\,r^2}\,.
\label{impartr+a}
\end{eqnarray}
Summing up Eqs. \eqref{finaltranssym} and \eqref{impartr+a} we obtain the complete contribution of the transverse modes
\begin{eqnarray}
\nonumber\delta \Sigma_s^{\rm (trans)}(E)&=&
-i\frac{2}{3}\als\,\cf T r^i(E-h_o)^2r^i-i\frac{\als \, C_F \, }{3}r^i(E-h_o)^3r^i
+i\frac{\als\cf\,Tm_D^2\,r^2\, }{3}\\
&&\times\left(\ln2-\frac12\right)+\order{\als T m_D^4r^2/E^2,\als r^2E^4/T,\als \,E \,m_D^2\,r^2}.
\label{tottrans}
\end{eqnarray}

We remark that the contribution of the transverse modes at the
energy scale is imaginary and finite, in contrast with what happens 
at zero temperature, where it is real and UV divergent, 
the divergence cancelling the infrared divergences appearing in the static, 
$1/m$ and $1/m^2$ potentials at the scale $1/r$. 
This is related to the discussion made in the previous section
regarding the cancellation of the IR divergence in Eqs. \eqref{cubic}
and \eqref{cubicshift} and can be understood in the following way.
For $E\gg m_D$, the Hard Thermal Loop transverse propagator can be
expanded for small $m_D$, giving, at the zeroth order, $(\Delta_{\rm R} +
\Delta_{\rm A})/2=i\mathrm{P} [{1}/{(k_0^2-k^2)}]$ and $(\Delta_{\rm R} - \Delta_{\rm
A})=2\pi\,\mathrm{sgn}(k_0)\delta(k^2_0-k^2)$. When plugged in
Eq. \eqref{deftransE} we obtain Eq. \eqref{transverseleading}. Evaluated at
the binding energy scale, the vacuum part is UV divergent and can be read from
\cite{Kniehl:1999ud,Brambilla:1999xj}: 
\begin{eqnarray}
&&g^2C_F\frac{D-2}{D-1} r^i\mu^{4-D}\int\frac{\,d^Dk}{(2\pi)^D}
\frac{k_0^2}{E-h_o-k_0+i\eta}\frac{i}{k_0^2-k^2+i\eta}r^i
\nonumber \\
&&=\frac{\als\cf}{3\pi}r^i(E-h_o)^3
\left(\frac{2}{\epsilon}+2\ln\frac{-(E-h_o)-i\eta}{\mu}+\gamma_E-\frac53-\ln\pi\right)r^i\,,
\label{USvacuum}
\end{eqnarray}
where the logarithm of the energy gives rise to the so-called QCD Bethe logarithm in the spectrum 
\cite{Kniehl:1999ud,Kniehl:2002br}. On the other hand, the temperature-dependent part gives
\begin{eqnarray}
\nonumber&& 
g^2C_F\frac{D-2}{D-1} r^i\mu^{4-D}\int\frac{\,d^Dk}{(2\pi)^D}\frac{k^2_0}{E-h_o-k_0+i\eta}
\left(\frac{T}{\vert k_0\vert}-\frac12+\order{\frac kT}\right)2\pi \delta\left(k_0^2-k^2\right)r^i
\\
\nonumber
&&=-i\frac{2}{3}\als\,\cf T r^i(E-h_o)^2r^i-\frac{\als\cf}{3\pi}r^i(E-h_o)^3
\left(\frac{2}{\epsilon}+2\ln\frac{\vert E-h_o\vert}{\mu}\right.
\\
&&\hspace{6.5cm}
\left.-i\pi\mathrm{sgn}(E-h_o)+\gamma_E-\frac53-\ln\pi\right)r^i\,,
\label{USmatter}
\end{eqnarray}
where the term proportional to $r^i(E-h_o)^2r^i$ comes from the first
term in the expansion of the Bose--Einstein distribution and the one
proportional to $r^i(E-h_o)^3r^i$ comes instead from the second term
in that expansion, see (\ref{boseexp}). In the sum of Eqs. \eqref{USvacuum} and
\eqref{USmatter} the real parts, divergences included, cancel out and
the imaginary parts combine to give the two $m_D$-independent terms of
Eq. \eqref{tottrans}. This shows  that the binding energy scale 
contribution produces two opposite UV divergences. 
In terms of the two interpretation schemes discussed in the previous section, 
we may understand the cancellation of divergences in two possible ways.
In the first way, the vacuum divergence in Eq. \eqref{USvacuum}
cancels the IR divergences of the potentials, whereas the UV matter divergence 
in \eqref{USmatter} cancels the IR matter divergence from the scale $T$ 
in \eqref{cubic}. In the second way, we consider the real part of the potential 
in  $\mathrm{pNRQCD}_\mathrm{HTL}$ as finite, the IR divergences from the scales $1/r$ 
and $T$ cancelling each other, and  no UV divergences coming  from the
energy scale, which, as shown by Eq. \eqref{tottrans}, is indeed the case. 
We stress that the cancellation of the divergences between the vacuum and thermal parts in
Eqs. \eqref{USvacuum} and \eqref{USmatter} is due to the second term
in the low-momentum expansion of the Bose--Einstein distribution,
i.e. $-1/2$, which is known in thermal field theory to cause
cancellations with the vacuum contribution. Finally, we observe that an
analogous cancellation is also obtained in the Abelian case \cite{Escobedo:2008sy}.

In order to obtain the contribution to the width from Eq. \eqref{tottrans}, 
we need to evaluate $r^i(E-h_o)^2r^i$. 
We proceed as in the previous section and rewrite $(E-h_o)^2$ 
as $(E-h_s)^2-\{(E-h_s),\Delta V\}+\Delta V^2$. One then has
\begin{equation}
r^i(E-h_o)^2r^i =\left(\frac{\nc^2}{4}\als^2+\frac{2\nc\als}{mr}+\frac{4\bp^2}{m^2}\right)
+ ...\,,
\label{squareexp}
\end{equation}
where the dots stand for contributions that vanish on the physical state. The width thus reads
\begin{eqnarray}
\nonumber
\Gamma_{n,l}^{\rm (trans)}&=&
\frac{1}{3}N_c^2C_F\als^3T-\frac{16}{3m}C_F\als TE_n+\frac{4}{3}N_cC_F\als^2T\frac{2}{mn^2a_0}
\\
\nonumber&&
+\frac{2E_n\als^3}{3}\left\{\frac{4 C_F^3\delta_{l0} }{n}+N_c \cf^2\left(\frac{8}{n (2l+1)}-\frac{1}{n^2} 
- \frac{2\delta_{l0}}{ n }   \right)+\frac{2N_c^2C_F}{n(2l+1)}+\frac{N_c^3}{4}\right\}
\\
&&-\frac{\als\cf\, (\ln4-1)\,Tm_D^2}{3}\frac{a_0^2n^2}{2}[5n^2+1-3l(l+1)]\,,
\label{totaltranssym}
\end{eqnarray}
where the first line is the contribution from the term proportional to
$r^i(E-h_o)^2r^i$, the second line comes from the cubic term and has
been obtained using Eqs. \eqref{cubicexp} and \eqref{cubicshift}, and
the third line is the contribution from the last term in the first line of Eq. \eqref{tottrans}.

The leading contribution to Eq. \eqref{totaltranssym}  
is given by the first three terms, which are of the same size.
The first term comes from the static potential and agrees with the one calculated in \cite{Brambilla:2009cd}.
The second and third terms come from the kinetic energy;
the second one agrees with the one calculated in \cite{Escobedo:2008sy}.
This contribution to the thermal decay width originates from the possible break up 
of a quark-antiquark colour-singlet state into an unbound quark-antiquark colour-octet state: 
a process that is kinematically allowed only in a medium \cite{Brambilla:2009cd}.
Clearly, the singlet to octet break up is a different phenomenon with respect 
to the Landau damping, which, in the previous section, 
provided another source for the in medium thermal width.  
In the situation $E \gg m_D$, which is the situation of 
interest for this work, the singlet to octet break up provides the dominant contribution 
to the thermal width. Indeed, comparing the Landau-damping width \eqref{GammanTloop} with 
the singlet to octet break-up width \eqref{totaltranssym}, we see that 
the latter is larger than the former by a factor $(m\als^2/m_D)^2$.

\subsection{Longitudinal gluon contribution}
The contribution of the longitudinal gluons to Eq. \eqref{defleading} is
\begin{eqnarray}
\nonumber\delta \Sigma_s^{\rm (long)}(E)&=&	
- i g^2 \, C_F \, \frac{r^i}{D-1}
\mu^{4-D} \int \frac{d^Dk}{(2\pi)^D} \frac{i}{E-h_o-k_0 +i\eta}k^2 
\left[ \frac{ D_{00}^{\rm R}(k_0,k) + D_{00}^{\rm A}(k_0,k)}{2}\right.\\
&&\hspace{3.8cm}
\left.+ \left(\frac{1}{2} + n_{\rm B}(k_0)\right)
\left(D_{00}^{\rm R}(k_0,k) - D_{00}^{\rm A}(k_0,k)\right)\right]r^i\,,
\label{deflongE}
\end{eqnarray}
where $D_{00}^{\rm R,A}(k)$ is the HTL propagator in
\eqref{prophtllong}. The first term in square brackets,
i.e. $(D_{00}^{\rm R} + D_{00}^{\rm A})/2$, does not depend on the
Bose--Einstein distribution; therefore only the expansion in $m_D\ll E$, 
corresponding to $m_D\ll k_0,k$, is possible. We then have
$(D_{00}^{\rm R} + D_{00}^{\rm
A})/2= {i}/{k^2}+\order{m_D^2/k^4}$. The first term is the free 
propagator, which gives a scaleless integration, whereas the second
one can be shown to contribute at order $\als E m_D^2 r^2$, which is
smaller than $m\als^5$.

For what concerns the symmetric part of the
propagator, i.e. $(1/2 + n_{\rm B}(k_0))(D_{00}^{\rm R} - D_{00}^{\rm
A})$, it should be noted that the retarded and advanced propagators
depend on $k_0$ only through the HTL self-energy; therefore, imaginary
parts in their denominators can enter only through the logarithm
appearing in Eq. \eqref{prophtllong}. Hence, the symmetric propagator
is non-zero solely in the spacelike $k^2>k_0^2$ region, which is
related to the Landau-damping phenomenon. At leading order in the
expansions of the Bose--Einstein distribution and of the propagator for
$m_D^2/k^2\ll 1$,  we thus have 
\begin{equation}
\left(\frac{1}{2} + n_{\rm B}(k_0)\right)\left(D_{00}^{\rm R}(k_0,k) - D_{00}^{\rm A}(k_0,k)\right)
=\frac{2\pi T m_D^2}{k^5}\theta\left(k^2-k_0^2\right)+\order{m_D^2/k^4,\,Tm_D^4/k^7}.
\label{deltashtllong}
\end{equation}
The first term contributes to the spectrum at order $\als Tm_D^2r^2$, 
so further terms in Eq. \eqref{deltashtllong} are not needed (see footnote \ref{footpre}). 
We then have
\begin{eqnarray}
\nonumber\delta \Sigma_s^{\rm (long)}(E)&=&
g^2C_F\frac{2}{D-1}r^i\mu^{4-D}
\int\frac{\,d^Dk}{(2\pi)^D}\frac{k^2}{E-h_o-k_0+i\eta}\frac{Tm_D^2\pi}{k^5}\theta\left(k^2-k_0^2\right)r^i\\
\nonumber&=&\frac{\als C_FTm_D^2}{6}r^i\left[2\pi\,\mathrm{sgn}(E-h_o)
+i\left(\frac{2}{\epsilon}+\ln\frac{(E-h_o)^2}{\mu^2}+\gamma_E-\frac83-\ln\pi\right)\right]r^i.\\
&&\label{tothtllong}
\end{eqnarray}

Equation \eqref{tothtllong} translates into the following shift of the energy levels 
\begin{equation}
\label{longshiftE}
\delta E_{n,l}^{\rm (long)}=-\frac{\pi\als C_F\ Tm_D^2}{3}\frac{a_0^2n^2}{2}[5n^2+1-3l(l+1)].
\end{equation}
For what concerns the width, we observe that the divergence is of
ultraviolet origin and cancels the one in Eq. \eqref{GammanTloop}, 
yielding a finite width; some care is, 
however, required in the handling of the logarithms of the energy, which
give rise to an analogue of the Bethe logarithm. We have
\begin{eqnarray}
\nonumber
\Gamma_{n,l}^{\rm (long)}&=&
-\frac{\als\cf Tm_D^2}{3}\left(\frac{2}{\epsilon}
+\ln\frac{E_1^2}{\mu^2}+\gamma_E-\frac83-\ln\pi\right)\frac{a_0^2n^2}{2}[5n^2+1-3l(l+1)]\\
  &&+\frac{2\als\cf Tm_D^2}{3}\frac{C_F^2\als^2}{E_n^2}\,I_{n,l}\;,
\label{longwidhtE}
\end{eqnarray}
where $E_1=-m\cf^2\als^2/4$ is the energy of the ground state and
\begin{equation}
I_{n,l}=\frac{E_n^2}{C_F^2\als^2}\int\frac{\,d^3k}{(2\pi)^3}
\left\vert\langle n,l\vert\br\vert\bk\rangle\right\vert^2\ln\frac{E_1}{E_n-{k^2}/{m}}\,.
\label{defin}
\end{equation}
$\langle n,l\vert\br\vert\bk\rangle$ is the matrix element between a
(bound) eigenstate $\vert n,l\rangle$ of $h_s$ and a continuum eigenstate
$\vert \bk\rangle$ of $h_o$.  This expression can be reduced to a single
integral using the techniques of \cite{Kniehl:1999ud,Kniehl:2002br}. 
We obtain for a singlet $nS$ state and an octet $P$ wave (the matrix element introduces a $\Delta l=1$
selection rule)
\begin{equation}
I_{n,0}=\int_0^\infty d\nu\, Y_n^{m_D}(\nu)X^2_n(\nu)\,,
\label{explin}
\end{equation}
where
\begin{equation}
Y_n^{m_D}(\nu)=\frac{\nu^6}{(\nu^2+\rho_n^2)^3}Y_n^E\,.
\label{Ynmd}
\end{equation}
The definitions of $Y_n^E$, $X^2_n$ for $n=1,2,3$ and 
$\rho_n$ can be found in \cite{Kniehl:1999ud} and
\cite{Kniehl:2002br}, the latter reference correcting some misprints
in the former. A numerical evaluation of these integrals for the three
most tightly bound $l=0$ states yields:
\begin{equation}
I_{1,0}=-0.49673,\qquad I_{2,0}=0.64070,\qquad I_{3,0}=1.18970.		
\label{numin}
\end{equation}

\subsection{Summary}
In summary, the contribution to the energy levels coming from the binding energy scale
is entirely due to the longitudinal part of the chromoelectric correlator, 
\begin{equation}
\label{totalspectrumE}
\delta E_{n,l}^{(E)}= \delta E_{n,l}^{\rm (long)},
\end{equation}
which may be read from Eq. \eqref{longshiftE}.
The contribution to the decay width coming from the binding energy scale
is the sum of $\Gamma_{n,l}^{\rm (trans)}$ and $\Gamma_{n,l}^{\rm (long)}$:
\begin{eqnarray}
\nonumber
\Gamma_{n,l}^{(E)}&=&
\frac{1}{3}N_c^2C_F\als^3T-\frac{16}{3m}C_F\als TE_n+\frac{8}{3}N_cC_F\als^2T\frac{1}{mn^2a_0}\\
\nonumber&&
+\frac{2E_n\als^3}{3}\left\{\frac{4 C_F^3\delta_{l0} }{n}+N_c \cf^2
\left(\frac{8}{n (2l+1)}-\frac{1}{n^2} - \frac{2\delta_{l0}}{ n }   \right)
+\frac{2N_c^2C_F}{n(2l+1)}+\frac{N_c^3}{4}\right\}\\
\nonumber&&
-\frac{\als\cf Tm_D^2}{6}\left(\frac{2}{\epsilon}
+\ln\frac{E_1^2}{\mu^2}+\gamma_E-\frac{11}{3}-\ln\pi+\ln4\right) a_0^2n^2 [5n^2+1-3l(l+1)]\\
 &&+\frac{2\als\cf Tm_D^2}{3}\frac{C_F^2\als^2}{E_n^2}\,I_{n,l}\;,
\label{totalwidthE}
\end{eqnarray}
where the first two lines come from the first two in
Eq. \eqref{totaltranssym} and the last two from Eq. \eqref{longwidhtE}
and from the last term in \eqref{totaltranssym}. $I_{n,l}$ is defined in
Eq. \eqref{defin}.

\section{Contributions to the spectrum from the scale $m_D$}
\label{secmD}
In our hierarchy of energy scales, the next scale after the binding energy 
is the Debye mass. 
We thus have to evaluate Eqs. \eqref{deftransE} and
\eqref{deflongE} for momenta of the order of $m_D$. In detail, we have
two regions to analyze: the first one is $k_0\sim E-h_o$, $k\sim m_D$,
corresponding to having the octet propagator unexpanded and conversely
expanding the HTL propagators for $k_0\gg k$. It can be easily shown
that both the transverse and the longitudinal parts result in a series
of scaleless integrations over $k$, which vanish in dimensional regularization.

The second region corresponds to having $k_0\sim
m_D$ and $k\sim m_D$: the octet propagator then needs to be expanded,
whereas the HTL propagators are to be kept in their resummed form. The
resulting integrals are quite involved, however, by power counting
arguments, it can be easily seen from Eqs. \eqref{deftransE} and
\eqref{deflongE} that, once the octet propagator is expanded, the
largest term comes again from the symmetric part of the gluon
propagator, due to the $T/k_0$ enhancement factor. The size of this
term turns out to be of order $\als T m_D^3r^2/E$ and, 
since we have assumed $\left({m_D}/{E}\right)^4\ll g$, it is beyond $m\als^5$ .

\section{Conclusions}
\label{secconclusions}
We have computed the heavy quarkonium energy levels and widths in a quark-gluon
plasma of temperature $T$ such that $m\als \gg  T \gg  m \als^2 \gg m_D$.
Assuming $(m_D/E)^4\ll g$, the spectrum is accurate up to order $m\als^5$. 

The thermal shift of the energy levels induced by the medium is obtained 
by summing the contribution from the scale $T$, given in
Eq. \eqref{totalenergyT}, with the thermal part of the contribution from 
the energy scale. We remark that the contribution from the energy scale, 
given in Eq. \eqref{totalspectrumE},  is the sum of both vacuum and thermal contributions, 
which, in the transverse sector, cancel. The thermal contribution of the transverse modes 
can be derived from  Eq. \eqref{USmatter}. 
The complete thermal contribution to the spectrum up to order $m\als^5$ reads
(we recall that $E_n=-{mC_F^2\als^2}/{(4n^2)}$ and $a_0 = {2}/{(m\cf\als)}$)
\begin{eqnarray}
\nonumber
\delta E_{n,l}^{(\mathrm{thermal})}&=&
\frac{\pi}{9}N_c C_F \,\als^2 \,T^2 \frac{a_0}{2}\left[3n^2-l(l+1)\right] +\frac{\pi}{3}C_F^2\, \als^2\, T^2\,a_0
\\
\nonumber&&
+\frac{E_n\als^3}{3\pi}\left[\log\left(\frac{2\pi T}{E_1}\right)^2-2\gamma_E\right]
\left\{\frac{4 C_F^3\delta_{l0} }{n}
+N_c \cf^2\left[\frac{8}{n (2l+1)}-\frac{1}{n^2} 
- \frac{2\delta_{l0}}{ n }   \right]\right.
\\
\nonumber 
&& \hspace{7cm}
\left. +\frac{2N_c^2C_F}{n(2l+1)} +\frac{N_c^3}{4}\right\}
\\
\nonumber
&&
+\frac{2E_n\cf^3\als^3}{3\pi}L_{n,l}
\\
\nonumber
&&
+ \frac{a_0^2n^2}{2}\left[5n^2+1-3l(l+1)\right]
\left\{- \left[\frac{3}{2\pi} \zeta(3)+\frac{\pi}{3}\right]  C_F \, \als  \, T \,m_D^2
\right.
\\
&& \hspace{7cm}
+ \left. \frac{2}{3} \zeta(3)\, N_c C_F \, \als^2 \, T^3\right\},
\label{finalspectrum}
\end{eqnarray}
where $L_{n,l}$ is the QCD Bethe logarithm, defined as \cite{Kniehl:1999ud,Kniehl:2002br} 
\begin{equation}
L_{n,l}=\frac{1}{C_F^2\als^2E_n}\int\frac{d^3k}{(2\pi)^3}
\left\vert\langle n,l\vert\br\vert\bk\rangle\right\vert^2\left(E_n-\frac{k^2}{m}\right)^3\ln\frac{E_1}{E_n-{k^2}/{m}}.
\label{defln}
\end{equation}
We refer to \cite{Kniehl:1999ud,Kniehl:2002br} for details on the
numerical evaluation of this integral. We furthermore remark that the
thermal contribution to the spectrum is finite, the IR divergence in
Eq. \eqref{totalenergyT} having cancelled against the UV divergence
coming from Eq. \eqref{USmatter}.

The thermal width is obtained by summing the
contribution from the scale $T$, given in Eq. \eqref{totalwidthT},
with the one coming from the energy scale as given in \eqref{totalwidthE}, 
the IR divergence in the former cancelling against the UV divergence in the latter. 
We then have
\begin{eqnarray}
\nonumber
\Gamma_{n,l}^{(\mathrm{thermal})}&=&
\frac{1}{3}N_c^2C_F\als^3T+\frac{4}{3}\frac{C_F^2\als^3 T}{n^2}(\cf+\nc)
\\
\nonumber&&
+\frac{2E_n\als^3}{3}\left\{\frac{4 C_F^3\delta_{l0} }{n}+N_c \cf^2
\left[\frac{8}{n (2l+1)}-\frac{1}{n^2} - \frac{2\delta_{l0}}{ n }   \right]
+\frac{2N_c^2C_F}{n(2l+1)}+\frac{N_c^3}{4}\right\}
\\
\nonumber&&
-\left[\frac{C_F}{6} \als T m_D^2
\left(\ln\frac{E_1^2}{T^2}+ 2\gamma_E -3 -\log 4- 2 \frac{\zeta^\prime(2)}{\zeta(2)} \right)
+\frac{4\pi}{9} \ln 2 \; N_c C_F \,  \als^2\, T^3 \right] 
\\
\nonumber
&& \hspace{8cm}\times\; a_0^2n^2\left[5n^2+1-3l(l+1)\right]
\\
&&
+\frac{8}{3}\cf\als\, Tm_D^2\,a_0^2n^4
\,I_{n,l}\;,
\label{finalwidth}
\end{eqnarray}
where $I_{n,l}$ is defined in Eq. \eqref{defin}. We remark that, up to the
order considered here, the thermal contribution to the spectrum and to
the width is independent of the spin.

Our results are expected to be relevant for the ground states of
bottomonium ($\Upsilon (1S)$ and $\eta_b$), and to a lesser extent to
those of charmonium ($J/\psi$ and $\eta_c$), for a certain range of
temperatures in the quark-gluon plasma for which (\ref{hierarchy}) is fulfilled. 
Let us now try to figure out what our results imply for 
the electromagnetic decays to lepton pairs or to two photons. First of all, the masses of
the heavy quarkonium states increase quadratically with the
temperature at leading order (first line of (\ref{finalspectrum})),
which would translate into the same functional increase in the
energy of the outgoing leptons and photons if produced by the quarkonium in the plasma. 
Second, since electromagnetic decays occur at short distances ($\sim 1/m \ll 1/T$), 
the standard NRQCD factorization formulas hold, and, at leading order, 
all the temperature dependence is encoded in the wave function
at the origin. The leading temperature correction to it comes from
first-order quantum-mechanical perturbation theory of the first term
of (\ref{totalpotT}). The size of this correction is $\sim n^4 T^2/(m^2\als)$. 
Hence, a quadratic dependence on the temperature should
also be observed in the frequency in which leptons or photons are
produced by the quarkonium in the plasma.
Finally, at leading order, a decay width linear with
temperature is developed (first line of (\ref{finalwidth})), which
implies a tendency to decay to the continuum of colour-octet states.
Hence, a smaller number of vector and pseudoscalar ground states is
expected to be in the sample with respect to the zero temperature case.

\acknowledgments
We acknowledge financial support from the RTN Flavianet MRTN-CT-2006-035482 (EU).  
N.B., J.G. and A.V  acknowledge financial support 
from the DFG cluster of excellence ``Origin and structure of the universe'' 
(\href{http://www.universe-cluster.de}{www.universe-cluster.de}). 
M.A.E. and J.S. acknowledges financial 
support from the FPA2007-60275/MEC grant (Spain) and the 2009SGR502 CUR grant (Catalonia). 
J.S also acknowledges financial support form the ECRI HadronPhysics2 
(Grant Agreement n. 227431) (EU), the FPA2007-66665-C02-01/MEC grant, 
and the Consolider Ingenio program CPAN CSD2007-00042 (Spain). 
M.A.E. has also been supported by a MEC FPU fellowship (Spain).

\appendix

\section{Details on the evaluation of the transverse HTL contribution}
\label{app_trans}
Our aim is the evaluation of Eq. \eqref{deftranssym}. Owing to the
symmetries of the retarded and advanced propagators and of the
Bose--Einstein distribution we can restrict the integration in
\eqref{deftranssym} to positive values of $k_0$. We then have
\begin{eqnarray}
\nonumber\delta \Sigma_s^{\rm (trans,\, symm)}(E)&=&g^2C_F\frac{D-2}{D-1}r^i\mu^{4-D}
\int\frac{\,d^{D-1}k}{(2\pi)^{D-1}}\int_0^\infty\frac{\,dk_0k_0^2}{2\pi} \left(\frac{T}{k_0}+\order{\frac ET}\right)\\
&& \hspace{-2.5cm}
\times\left(\Delta_{\rm R}(k_0,k) 
- \Delta_{\rm A}(k_0,k)\right)\left(\frac{1}{E-h_o-k_0+i\eta}+\frac{1}{E-h_o+k_0+i\eta}\right)r_i\,.
\label{pot}
\end{eqnarray}

Let us define the quantity $\lambda\equiv k_0-k$.
There exist two momentum regions 
that contribute to the integral \eqref{pot} for $k_0\sim k\sim E-h_o$.
We call the first region the \emph{off-shell region}. It is defined by
\begin{equation}
\label{defhard}
\lambda\sim (E-h_o)\,,\qquad k\sim (E-h_o)\,,
\end{equation}
i.e. the region where the gluon is far from being on shell.
The second region is called the \emph{collinear region}. In this region, we have
\begin{equation}
\label{defcoll}
\lambda\sim\frac{m_D^2}{E-h_o}\,,\qquad k\sim (E-h_o)\,.
\end{equation}
We observe that the collinear scale $m_D^2/(E-h_o)$ has,
in our energy scale hierarchy, a magnitude in between $mg^4$ and $mg^6$. 
It is, therefore, smaller that the Debye
mass by a factor of $m_D/E\ll 1$ and still larger than the
non-perturbative magnetic mass, which is of order $g^2 T$, by a factor $T/E\gg1$.
For simplicity, we separate the two regions by a cut-off $\Lambda$, such that
\begin{equation}
(E-h_o)\gg \Lambda\gg\frac{m_D^2}{(E-h_o)}\,.
\end{equation}

We start by analyzing the off-shell region. Here $k_0^2-k^2=\lambda(2k+\lambda)\gg m_D^2$ 
and we can thus expand the retarded propagator propagator in Eq. \eqref{prophtltrans} as
\begin{equation}
\Delta_\mathrm{R}(k_0>0)=
\frac{i}{k_0^2-k^2+i\eta}+\frac{i\frac{m_D^2}{2}\left(\frac{k_0^2}{k^2}-(k_0^2-k^2)
\frac{k_0}{2k^3}\ln\left(\frac{k_0+k+i\eta}{k_0-k+i\eta}\right)\right)}{(k_0^2-k^2+i\eta)^2}
+\mathcal{O}\left(\frac{m_D^4}{(E-h_o)^6}\right).
\end{equation}
Terms contributing to the real part of this propagator and hence to
$\Delta_\mathrm{R}-\Delta_\mathrm{A}$ can come either from the poles
of the denominators, yielding a $\delta(k_0^2-k^2)$, or from the
imaginary part of the logarithm. However, $\delta(k_0^2-k^2)=0$
over the whole off-shell region. We can safely discard these terms and obtain
\begin{equation}
(\Delta_\mathrm{R}-\Delta_{A})(k_0>0)=-\frac{m_D^2k_0\pi\theta(k-k_0)}{2k^3}\mathrm{P}\frac{1}{k_0^2-k^2}.
\end{equation}
Note that the principal value prescription is irrelevant since our integration region excludes the poles. 
From Eq. \eqref{pot}, we get
\begin{eqnarray}
\nonumber
\delta \Sigma^{\rm (trans,\, symm)}_{s,\mathrm{off \; shell}}(E)&=&
-\frac{g^2C_Fm_D^2T\pi r^i(D-2)}{2(D-1)}\int\frac{\,d^{D-1}k}{(2\pi)^{D-1}}\frac{1}{k^3}
\\
&&\times\int_0^{k-\Lambda}\! \frac{dk_0}{2\pi}\mathrm{P}\frac{k_0^2}{k_0^2-k^2}
\frac{2(E-h_o+i\eta)}{(E-h_o+i\eta)^2-k_0^2}r^i\,.
\end{eqnarray}
This integral does not need to be dimensionally regularized, 
so we can set $D=4$ at this point and obtain 
\begin{equation}
\delta \Sigma^{\rm (trans,\, symm)}_{s,\mathrm{off \; shell}}(E)=-\frac{g^2C_Fm_D^2T r^i}{12\pi}
\int_0^\infty\!\frac{dk_0}{2\pi}\frac{2(E-h_o+i\eta)}{(E-h_o+i\eta)^2-k_0^2}
\left[\ln\frac{2\Lambda}{k_0}+\mathcal{O}\left(\frac{\Lambda}{k_0}\right)\right]r^i\,.
\label{hardtrans}
\end{equation}

We consider, now, the collinear region.
We start again from the retarded propagator introduced in
Eq. \eqref{prophtltrans}. We perform the change of variables
$k_0-k=\lambda$ and we expand for $\lambda\sim {m_D^2}/{k}\ll k$,
thereby implementing the collinear hierarchy. We then have
\begin{equation}
(\Delta_\mathrm{R}-\Delta_\mathrm{A})(k_0>0)=\Delta_1+\Delta_2+\Delta_3+\Delta_4+\Delta_5+\order{\frac{\lambda}{k^3}},
\end{equation}
where the $\Delta_i$ are defined as
\begin{equation}
\Delta_1=\frac{i}{2k}\left(\frac{1}{\lambda-\frac{m_D^2}{4k}+i\eta}-\frac{1}{\lambda-\frac{m_D^2}{4k}-i\eta}\right),
\end{equation}
\begin{equation}
\Delta_2=\frac{3im_D^4}{64k^4}
\left(\frac{1}{(\lambda-\frac{m_D^2}{4k}+i\eta)^2}-\frac{1}{(\lambda-\frac{m_D^2}{4k}-i\eta)^2}\right),
\end{equation}
\begin{equation}
\Delta_3=-\frac{im_D^2}{8k^3}\left(\frac{\ln\left(\frac{2k}{\lambda+i\eta}\right)}
{\lambda-\frac{m_D^2}{4k}+i\eta}-\frac{\ln\left(\frac{2k}{\lambda-i\eta}\right)}{\lambda-\frac{m_D^2}{4k}-i\eta}\right),
\end{equation}
\begin{equation}
\Delta_4=-\frac{im_D^4}{32k^4}\left(\frac{\ln\left(\frac{2k}{\lambda+i\eta}\right)}
{(\lambda-\frac{m_D^2}{4k}+i\eta)^2}
-\frac{\ln\left(\frac{2k}{\lambda-i\eta}\right)}{(\lambda-\frac{m_D^2}{4k}-i\eta)^2}\right),
\end{equation}
\begin{equation}
\Delta_5=\frac{im_D^2}{8k^3}\left(\frac{1}{\lambda-\frac{m_D^2}{4k}+i\eta}
-\frac{1}{\lambda-\frac{m_D^2}{4k}-i\eta}\right).
\end{equation}
We start by plugging $\Delta_1$ in Eq. \eqref{pot}. We then have 
\begin{eqnarray}
\nonumber
\delta \Sigma^{\rm (trans,\, symm)}_{s,1}(E)&=&
\frac{g^2C_F}{6}r^i\int\frac{\,d^3k}{(2\pi)^3}\int_{-\Lambda}^{\Lambda}\! d\lambda(k+\lambda)\frac{T}{k}
\delta\left(\lambda-\frac{m_D^2}{4k}\right)
\\
&& \hspace{5cm}
\times \frac{2(E-h_o+i\eta)}{(E-h_o+i\eta)^2-(k+\lambda)^2}r^i
\nonumber
\\
&=&-i\frac{2}{3}\als\,\cf\,T\,r^i(E-h_o)^2r^i+\order{\frac{\als T m_D^4r^2}{E^2}}.
\label{tot1}
\end{eqnarray}
The contribution of $\Delta_2$ is 
\begin{eqnarray*}
\delta \Sigma^{\rm (trans,\, symm)}_{s,2}(E)&=&\frac{g^2C_FTr^i}{3}\int\frac{\,d^3k}{(2\pi)^3}
\int_{-\Lambda}^\Lambda\frac{\,d\lambda}{2\pi}\frac{3im_D^4}{32k^3}
\\
&& \times 
\left(\frac{1}{(\lambda-\frac{m_D^2}{4k}+i\eta)^2}-\frac{1}{(\lambda-\frac{m_D^2}{4k}-i\eta)^2}\right) 
\frac{2(E-h_o+i\eta)}{(E-h_o+i\eta)^2-(k+\lambda)^2}r^i\\
&=&\order{\als T m_D^4r^2/E^2};
\end{eqnarray*}
the leading order term in the expansion of $((E-h_o+i\eta)^2-(k+\lambda)^2)^{-1}$, 
which would contribute at order $\als T m_D^2 r^2$, vanishes because the integral over $\lambda$ is zero.
The contribution of $\Delta_3$ is 
\begin{eqnarray*}
\delta \Sigma^{\rm (trans,\, symm)}_{s,3}(E)&=&
-\frac{ig^2C_FTm_D^2r^i}{12}\int\frac{\,d^3k}{(2\pi)^3}
\frac{1}{k^2} \int_{-\Lambda}^\Lambda\frac{\,d\lambda}{2\pi}
\frac{2(E-h_o+i\eta)}{(E-h_o+i\eta)^2-(k+\lambda)^2} \\
&&\times
\left[\ln\left\vert\frac{2k}{\lambda}\right\vert\left(-2i\pi\delta
\left(\lambda-\frac{m_D^2}{4k}\right)\right)
-2i\pi\theta(-\lambda)\mathrm{P}\frac{1}{\lambda-\frac{m_D^2}{4k}}\right]r^i\,.
\end{eqnarray*}
We then have
\begin{equation}
\delta \Sigma^{\rm (trans,\, symm)}_{s,3}(E)=
-\frac{g^2 C_F Tm_D^2r^i}{12\pi}\int_0^\infty\! \frac{dk}{2\pi} \frac{2(E-h_o+i\eta)}{(E-h_o+i\eta)^2-k^2}
\left[\ln\left(\frac{2k}{\Lambda}\right)+...\right]r^i\,,
\end{equation}
where the dots mean terms suppressed by ${1}/{\Lambda}$. 
We now combine this result with the contribution from the off-shell region in Eq. \eqref{hardtrans} to obtain
\begin{eqnarray}
\nonumber\delta \Sigma^{\rm (trans,\, symm)}_{s,\mathrm{off \; shell}}(E)+\delta \Sigma^{\rm (trans,\, symm)}_{s,3}(E)&=&
-\frac{\als \cf Tm_D^2r^i}{6\pi}\ln4\int_0^\infty\! dk\frac{2(E-h_o+i\eta)}{(E-h_o+i\eta)^2-k^2}r^i \\
&=&i\frac{\als \cf Tm_D^2r^2}{3}\ln2+\ldots\,,
\label{c+hard}
\end{eqnarray}
where the dots stand for higher orders. We remark that the dependence on 
the cut-off scale $\Lambda$ has disappeared.
The contribution of $\Delta_4$ is
\begin{eqnarray*}
&& \delta \Sigma^{\rm (trans,\, symm)}_{s,4}(E)=
-\frac{ig^2Tm_D^4C_Fr^i}{48}\int\frac{\,d^3k}{(2\pi)^3}\frac{1}{k^3}
\int_{-\Lambda}^\Lambda\frac{\,d\lambda}{2\pi}\frac{2(E-h_o+i\eta)}{(E-h_o+i\eta)^2-(k+\lambda)^2}
\\
&& \hspace{3cm}  \times 
\left[\left(\frac{\ln\left\vert\frac{2k}{\lambda}\right\vert}
{(\lambda-\frac{m_D^2}{4k}+i\eta)^2}-\frac{\ln\left\vert\frac{2k}{\lambda}\right\vert}
{(\lambda-\frac{m_D^2}{4k}-i\eta)^2}\right)
-\frac{2i\pi\theta(-\lambda)}{(\lambda-\frac{m_D^2}{4k}-i\eta)^2}\right]r^i\,.
\end{eqnarray*}
The needed $\lambda$ integrals are 
\begin{eqnarray*}
\int_{-\Lambda}^\Lambda\frac{\,d\lambda}{2\pi}
\ln\left\vert\frac{2k}{\lambda}\right\vert\left(\frac{1}{(\lambda-\frac{m_D^2}{4k}+i\eta)^2}
-\frac{1}{(\lambda-\frac{m_D^2}{4k}-i\eta)^2}\right)=i\frac{4k}{m_D^2}\,,
\end{eqnarray*}
and
\begin{equation*}
-i\int_{-\Lambda}^\Lambda\,d\lambda\frac{\theta(-\lambda)}{(\lambda-\frac{m_D^2}{4k}-i\eta)^2}=
-\frac{i4k}{m_D^2}+...\,,
\end{equation*}
so that $\delta \Sigma^{\rm (trans,\, symm)}_{s,4}(E)$ has only contributions that are suppressed 
by powers of ${1}/{\Lambda}$.

Finally, the contribution of $\Delta_5$ is 
\begin{eqnarray}
\nonumber
\delta \Sigma^{\rm (trans,\, symm)}_{s,5}(E)&=&
\frac{g^2C_Fr^i(D-2)}{2(D-1)}
\int\frac{\,d^{D-1}k}{(2\pi)^{D-1}}\int_{-\Lambda}^\Lambda\frac{\,d\lambda}{2\pi}
\frac{T\pi m_D^2}{2k^2}\delta(\lambda-\frac{m_D^2}{4k})
\\
\nonumber
&& \hspace{6cm}
\times \frac{2(E-h_o+i\eta)}{(E-h_o+i\eta)^2-(k+\lambda)^2}r^i 
\\
&=&-i\frac{\als \cf Tm_D^2r^2}{6} +\ldots\,,
\label{finale}
\end{eqnarray}
where the dots stand for higher orders.
The contribution 
of the symmetric part of the transverse propagator is then given 
by the sum of Eqs. \eqref{tot1}, \eqref{c+hard} and \eqref{finale}.


\begin{thebibliography}{99}

\bibitem{Matsui:1986dk}
  T.~Matsui and H.~Satz,
  Phys.\ Lett.\  B {\bf 178}, 416 (1986).

\bibitem{Lourenco:2006sr}
  C.~Louren\c{c}o,
  Nucl.\ Phys.\  A {\bf 783}, 451 (2007) 
  [arXiv:nucl-ex/0612014].

\bibitem{Laine:2006ns}
  M.~Laine, O.~Philipsen, P.~Romatschke and M.~Tassler,
  JHEP {\bf 0703}, 054 (2007)
  [arXiv:hep-ph/0611300].

\bibitem{Laine:2007gj}
  M.~Laine,
  JHEP {\bf 0705}, 028 (2007)
  [arXiv:0704.1720 [hep-ph]].

\bibitem{Burnier:2007qm}
  Y.~Burnier, M.~Laine and M.~Vepsalainen,
  JHEP {\bf 0801}, 043 (2008)
  [arXiv:0711.1743 [hep-ph]].

\bibitem{Beraudo:2007ky}
  A.~Beraudo, J.~P.~Blaizot and C.~Ratti,
  Nucl.\ Phys.\  A {\bf 806}, 312 (2008)
  [arXiv:0712.4394 [nucl-th]].

\bibitem{Escobedo:2008sy}
  M.~A.~Escobedo and J.~Soto,
  Phys. Rev. A {\bf 78}, 032520 (2008), [arXiv:0804.0691 [hep-ph]].

\bibitem{Brambilla:2009cd}
  N.~Brambilla, J.~Ghiglieri, A.~Vairo and P.~Petreczky,
  Phys.\ Rev.\  D {\bf 78}, 014017 (2008) 
  [arXiv:0804.0993 [hep-ph]].

\bibitem{Petreczky:2005bd}
  P.~Petreczky,
  Eur.\ Phys.\ J.\  C {\bf 43}, 51 (2005)
  [arXiv:hep-lat/0502008].

\bibitem{Satz:2005hx}
  H.~Satz,
  J.\ Phys.\ G {\bf 32}, R25 (2006)
  [arXiv:hep-ph/0512217].

\bibitem{Laine:2008cf}
  M.~Laine,
  Nucl.\ Phys.\  A {\bf 820}, 25C (2009)
  [arXiv:0810.1112 [hep-ph]].

\bibitem{Brambilla:2004jw}
  N.~Brambilla, A.~Pineda, J.~Soto and A.~Vairo,
  Rev.\ Mod.\ Phys.\  {\bf 77}, 1423 (2005) 
  [arXiv:hep-ph/0410047].

\bibitem{muonic}
  M.~A.~Escobedo and J.~Soto, UB-ECM-PF 09/15, in preparation.

\bibitem{Landshoff:1992ne}
  P.~V.~Landshoff and A.~Rebhan,
  Nucl.\ Phys.\  B {\bf 383}, 607 (1992)
  [Erratum-ibid.\  B {\bf 406}, 517 (1993)]
  [arXiv:hep-ph/9205235].

\bibitem{Caswell:1985ui}
  W.~E.~Caswell and G.~P.~Lepage,
  Phys.\ Lett.\ B {\bf 167}, 437 (1986);
  G.~T.~Bodwin, E.~Braaten and G.~P.~Lepage,
  Phys.\ Rev.\ D {\bf 51}, 1125 (1995)
  [Erratum-ibid.\ D {\bf 55}, 5853 (1997)]
  [hep-ph/9407339].

\bibitem{Pineda:1997bj}
  A.~Pineda and J.~Soto,
  Nucl.\ Phys.\ Proc.\ Suppl.\  {\bf 64}, 428 (1998)
  [arXiv:hep-ph/9707481].

\bibitem{Brambilla:1999xf}
  N.~Brambilla, A.~Pineda, J.~Soto and A.~Vairo,
  Nucl.\ Phys.\  B {\bf 566}, 275 (2000)
  [arXiv:hep-ph/9907240].

\bibitem{Smirnov:2009fh}
  A.~V.~Smirnov, V.~A.~Smirnov and M.~Steinhauser,
  Phys.\ Rev.\ Lett.\  {\bf 104}, 112002 (2010)
  [arXiv:0911.4742 [hep-ph]].

\bibitem{Anzai:2009tm}
  C.~Anzai, Y.~Kiyo and Y.~Sumino,
  Phys.\ Rev.\ Lett.\  {\bf 104}, 112003 (2010)
  [arXiv:0911.4335 [hep-ph]].

\bibitem{Kniehl:2004rk}
  B.~A.~Kniehl, A.~A.~Penin, Y.~Schr\"oder, V.~A.~Smirnov and M.~Steinhauser,
  Phys.\ Lett.\  B {\bf 607}, 96 (2005) 
  [arXiv:hep-ph/0412083].

\bibitem{Appelquist:1977es}
  T.~Appelquist, M.~Dine and I.~J.~Muzinich,
  Phys.\ Rev.\  D {\bf 17}, 2074 (1978).

\bibitem{Brambilla:1999qa}
  N.~Brambilla, A.~Pineda, J.~Soto and A.~Vairo,
  Phys.\ Rev.\  D {\bf 60}, 091502 (1999)
  [arXiv:hep-ph/9903355].

\bibitem{Brambilla:1999xj}
  N.~Brambilla, A.~Pineda, J.~Soto and A.~Vairo,
  Phys.\ Lett.\  B {\bf 470}, 215 (1999)
  [arXiv:hep-ph/9910238].

\bibitem{Kniehl:2002br}
  B.~A.~Kniehl, A.~A.~Penin, V.~A.~Smirnov and M.~Steinhauser,
  Nucl.\ Phys.\  B {\bf 635}, 357 (2002) 
  [arXiv:hep-ph/0203166].

\bibitem{Penin:2002zv}
  A.~A.~Penin and M.~Steinhauser,
  Phys.\ Lett.\  B {\bf 538} (2002) 335
  [arXiv:hep-ph/0204290].

\bibitem{Vairo:2009ih}
  A.~Vairo,
  PoS {\bf CONFINEMENT8}, 002 (2008)
  [arXiv:0901.3495 [hep-ph]].

\bibitem{Braaten:1991gm}
  E.~Braaten and R.~D.~Pisarski,
  Phys.\ Rev.\  D {\bf 45}, 1827 (1992).

\bibitem{Kniehl:1999ud}
  B.~A.~Kniehl and A.~A.~Penin,
  Nucl.\ Phys.\  B {\bf 563} (1999) 200
  [arXiv:hep-ph/9907489].

\bibitem{Titard:1993nn}
  S.~Titard and F.~J.~Yndurain,
  Phys.\ Rev.\  D {\bf 49}, 6007 (1994)
  [arXiv:hep-ph/9310236].

\bibitem{Brambilla:2003nt}
  N.~Brambilla, D.~Gromes and A.~Vairo,
  Phys.\ Lett.\  B {\bf 576}, 314 (2003)
  [arXiv:hep-ph/0306107].

\bibitem{Carrington:1997sq}
  M.~E.~Carrington, D.~f.~Hou and M.~H.~Thoma,
  Eur.\ Phys.\ J.\  C {\bf 7}, 347 (1999)
  [arXiv:hep-ph/9708363].

\end{thebibliography}
\end{document}